\documentclass{aa}  
\usepackage{graphicx}
\usepackage{txfonts}
\usepackage{float}
\begin{document} 

   \title{Spatially resolved spectroscopy across stellar surfaces. III. }

   \subtitle{Photospheric Fe I lines across HD~189733A (K1 V) }
  
   \author{Dainis Dravins
          \inst{1}, 
               Martin Gustavsson
        \inst{1},
               \and
      Hans-G\"{u}nter Ludwig
           \inst{2} 
                }
%
% Please retain full first names of authors! 
%
\institute{Lund Observatory, Box 43, SE-22100 Lund, Sweden\\
              \email{dainis@astro.lu.se}
\and
     Zentrum f\"{u}r Astronomie der Universit\"{a}t Heidelberg, Landessternwarte, K\"{o}nigstuhl, DE--69117 Heidelberg, Germany\\
              \email{hludwig@lsw.uni-heidelberg.de}
             }
 
\date{Received March 13, 2018; accepted May 15, 2018}

% \abstract{}{}{}{}{} 
% 5 {} token are mandatory
 
\abstract
   {Spectroscopy across spatially resolved stellar surfaces reveals spectral line profiles free from rotational broadening, whose gradual changes from disk center toward the stellar limb reflect an atmospheric fine structure that is possible to model by  3\mbox{-}D hydrodynamics.}
   {Previous studies of photospheric spectral lines across stellar disks exist for the Sun and HD\,209458 (G0 V) and are now extended to the planet-hosting HD\,189733A to sample a cooler K-type star and explore the future potential of the method. }
   {During exoplanet transit, stellar surface portions successively become hidden and differential spectroscopy between various transit phases uncovers spectra of small surface segments temporarily hidden behind the planet.  The method was elaborated in Paper~I, in which observable signatures were predicted quantitatively from hydrodynamic simulations.}
   {From observations of HD\,189733A with the ESO HARPS spectrometer at $\lambda/\Delta \lambda\sim$115,000, profiles for stronger and weaker \ion{Fe}{I} lines are retrieved at several center-to-limb positions, reaching adequate S/N after averaging over numerous similar lines.}
   {Retrieved line profile widths and depths are compared to synthetic ones from models with parameters bracketing those of the target star and are found to be consistent with 3-D simulations.  Center-to-limb changes strongly depend on the surface granulation structure and much greater line-width variation is predicted in hotter F-type stars with vigorous granulation than in cooler K-types.  Such parameters, obtained from fits to full line profiles, are realistic to retrieve for brighter planet-hosting stars, while their hydrodynamic modeling offers previously unexplored diagnostics for stellar atmospheric fine structure and 3\mbox{-}D line formation.  Precise modeling may be required in searches for Earth-analog exoplanets around K-type stars, whose more tranquil surface granulation and lower ensuing microvariability may enable such detections. }
% Comment: The wrtiting `naked' is retained since the intended meaning is "so-called naked" which is not the same as "naked".  

\titlerunning{Spatially resolved spectroscopy across stellar surfaces. III.}
\authorrunning{D. Dravins et al.}
\maketitle

\section{Introduction}

Three-dimensional and time-dependent hydrodynamic simulations provide realistic descriptions for the photospheres of the various classes of stars.    
The combination of stellar spectra computed from such models, with corresponding models of planetary atmospheres, constrain the properties of both the star and its exoplanets. To test and evolve such models, more detailed observations than the ordinary spectrum of integrated starlight are desirable.  Spectral line syntheses from 3\mbox{-}D atmospheres display numerous features in the gradual changes of photospheric line profile strengths, shapes, asymmetries, and wavelength shifts from the stellar disk center toward the stellar limb.  Direct comparisons between theory and spectral-line observations were in the past carried out only for the spatially resolved Sun \citep[e.g.,][]{lindetal17}. This project extends such studies to other stars as well.  In Paper~I \citep{dravinsetal17a}, theoretically predicted spatially resolved signatures were examined for a group of main-sequence stellar models with temperatures between 3960 and 6730\,K.  Corresponding observations are feasible during exoplanet transits when small stellar surface portions successively become hidden, and differential spectroscopy between different transit phases provide spectra of small surface segments temporarily hidden behind the planet.  The observational requirements were elaborated in Paper~I, in which observable signatures were predicted quantitatively.  In Paper~II \citep{dravinsetal17b}, a first observational demonstration of this method was presented for the G0~V star HD\,209458.  

This Paper~III extends such studies to another bright planet-hosting star but of lower temperature, the m$_{\textrm{V}}$ = 7.7, K1~V star HD\,189733A, circled by its planet HD\,189733Ab.  To avoid constantly repeating their (easily intermixed)  numbers from the Henry Draper catalog, for the purposes of this paper we refer to HD\,189733A as {\it{`Alopex'}}\footnote{This name -- Greek  $A\lambda\acute{\omega}\pi\eta\xi$ -- is taken from the original taxonomic name for the arctic fox ({\it{Alopex Lagopus}}) given by Linnaeus in 1758, here relating to the namesake of its constellation Vulpecula (Latin for `little fox').  It is also the Greek name for that constellation. }, and to HD\,189733Ab as simply `the planet'.  

Besides probing a cooler stellar atmosphere, there is a significance for exoplanet searches since the surfaces of K-type stars are predicted to be quieter, with lesser amplitudes in velocity and temperature than in hotter and more solar-like stars. 

Searches for exoplanets that mimic the Earth in terms of mass and orbit are limited by intrinsic stellar microvariability in either apparent radial velocity or photometric brightness \citep{fischeretal16}, and therefore may appear less challenging around K-type stars.  This requires adequate modeling of their surface convection so that, for example, correlations between stellar fluctuations in brightness, color, and radial velocity are adequately understood and may be compensated for. However, given the often greater magnetic activity and increased spottiness on such stars, separate treatments of magnetically sensitive features over the timescales of their variability may also be needed to disentangle exoplanet signatures. 

Further, analyses of exoplanet atmospheres by transmitted starlight require the background stellar spectrum to be precisely known.  That is not the averaged spectrum of the entire stellar disk but that from only the small area temporarily behind the planet, such that the varying stellar spectrum along the chord of transit has to be known for any more precise studies \citep{rackhametal18}.

Besides applying this method to a new target star, the aim is to explore its general practicality.  Currently, large transiting planets with bright host stars are actively being sought by several photometric surveys. Recent and forthcoming high-resolution spectrometers at major telescopes are likely to observe such systems for exoplanet research; using this method, the same data can be used to probe also the host star.

\section{Hydrodynamic atmospheres and their spectra}

\subsection{Three-dimensional structures on different stars}

Three-dimensional and time-variable model photospheres are now established as realistic descriptions of stellar outer layers.  Originating from initial modeling of the solar photosphere \citep[e.g.,][]{nordlundetal09}, simulations have now been extended to stars with widely different parameters over much of the Hertzsprung-Russel diagram \citep{magicetal13, tremblayetal13}.  Although the basic physical processes are similar in giants, dwarfs, and even white dwarfs, synthetic surface images reveal the greatly changing granulation sizes on stars with different surface gravity, as well as how the granulation brightness contrast increases with temperature, at least for stars comparable to the Sun.  In disk-integrated starlight, predicted observable quantities include photometric microvariability and flickering \citep{samadietal13a,samadietal13b}.  Signatures of the surface granulation might also be observable during exoplanet transits as an additional intensity fluctuation (with a certain correlation between different colors) caused by its temporal variability on spatial scales that are differently sampled by planets of dissimilar size \citep{chiavassaetal17}.  Other classes of predictions differing between classical 1\mbox{-}D and 3\mbox{-}D models are those of stellar limb darkening \citep{magicetal15} and of photometric colors \citep{kucinskasetal18}.

Of particular relevance for cool K-type stars such as Alopex, is that the granulation contrast is predicted to be rather lower than in somewhat hotter stars more similar to the Sun.  Not only does the amplitude of convection itself get reduced, but also the extent to which its effects are visible on the stellar surface.  The lower effective temperature of course implies a lower stellar flux but also decreases the atmospheric opacity at a given pressure, correspondingly increasing the characteristic densities near the surface.  Thus, the stellar surface (optical density $\tau\sim$1) is located at higher densities.  With higher densities and less convective flux, smaller velocity and temperature amplitudes suffice to transport the stellar flux outward. Further, the opacity structure enables radiative diffusion to carry a substantial fraction of the flux in the layers just below the visible surface, reducing the convective flux in the visible layers to only a small fraction of the total.  For modeling details and comparisons between synthetic granulation images for stars at different temperature, see \citet{magicetal13}, \citet{nordlunddravins90}, and \citet{tremblayetal13}.

\subsection{K-dwarf surface spectroscopy}

As stressed in Paper~I, the simulation of 3\mbox{-}D atmospheric hydrodynamics is only a first part of the problem since most astrophysical deductions about stellar or planetary properties require synthetic spectra for comparison; see, for example, \citet{asplundetal00}, \citet{asplund05} and \citet{chiavassaetal18}. Synthetic spectral lines from such K-type stellar models have been computed by \citet{dravinsnordlund90b} and \citet{ramirezetal09}.  The lower amplitudes of convective motion and lesser temperature contrast result in smaller convective wavelength blueshifts and weaker curvatures of stellar bisectors, as compared to, for instance, the solar case.  Some observations have been made to seek evidence of the surface structure on Alopex and other K-type dwarfs.  Thus, \citet{ramirezetal08} obtained high-resolution spectra of several K-type dwarfs to identify asymmetries in differently strong \ion{Fe}{I} line profiles that have C-shaped bisectors as signatures of granulation.  

Specifically for Alopex, observational confrontations with 3\mbox{-}D models have been carried out in analyzing photometric light curves during transit, searching for effects to segregate between limb-darkening functions predicted from 1\mbox{-}D and 3\mbox{-}D model atmospheres.  The differences are small, however, and largely masked by the great amount of absorption lines, whose different and complex center-to-limb behavior largely masks the underlying temperature structure \citep{hayeketal12}.  If extremely good photometric precision could be achieved, polarimetric observations during transit could offer further constraints on atmospheric structure, as simulated by \citet{kostogryzetal17}.

An ambitious study to deduce the center-to-limb variation of the strong absorption lines of \ion{Ca}{II}~H \& K and \ion{Na}{I}~D$_{\textrm{1}}$ \& D$_{\textrm{2}}$ lines was carried out by \citet{czeslaetal15}, recording transit spectra of Alopex with the ESO UVES spectrometer.  The wings of these lines were found to be limb-brightened relative to the spectral continuum, in agreement with models.  Understanding the behavior of such very strong lines is essential for exoplanet atmospheric studies, since these are the types of lines in which their atmospheres can be detected.  However, corresponding behavior at line centers could not be identified because this behavior was masked by variable stellar activity.

\section{HD\,189733A ({\textbf{\textit{Alopex}}}): The star and its planet}

Although the principle of differential spectroscopy during exoplanet transit may appear to be relatively straightforward, its practical implementation requires observations at very high signal-to-noise, as emphasized in Paper~I.  Transiting planets cover only a tiny fraction of the stellar surface, which currently limits the method to the brightest planet-hosting stars with the largest planets.  In this work, this method is applied to HD\,189733A (Alopex).  Its transiting planet covers {$\sim$}2.5 \% of the stellar surface and produces the deepest transit among the currently known brighter host stars (Fig.\ 1 in Paper~II).  We used photometry from the Hubble Space Telescope (HST) by \citet{pontetal07}, who measured the transit depth as 2.6\%  of the stellar irradiance (Fig.\ \ref{fig:planet_positions}).  The precise value, however, depends on the exact wavelength range, and its limb darkening at the chord of transit.   

Along with its apparent brightness and deep photometric transit, an advantage in choosing this target is that its planet HD\,189733Ab has, since its discovery \citep{bouchyetal05}, been intensely studied, and many data are available in various archives. The system remains a prime target for exoplanet atmospheric studies and is certain to be observed with future instrumentation.  
 
Numerous studies of the planet, including the dynamics of its extended atmosphere, have been made.  Daytime temperatures of $\sim$1200~K in line-forming regions \citep[e.g.,][]{knutsonetal07,lineetal16} imply a bloated atmosphere with layers of light gases and molecular constituents such as H$_{\textrm{2}}$O, CH$_{\textrm{4}}$, and CO$_{\textrm{2}}$.  While this inflated state enables the study of the planet's atmospheric chemistry and mass loss, the presence of spectral lines from such planetary species along with the different effective planet sizes in the corresponding monochromatic light imply that constructions of the observed (star+planet) spectra at the precise wavelengths corresponding to such molecular lines could be a more complex task;  we do not  attempt this approach in this work.  Rather, this study targets stellar photospheric \ion{Fe}{I} lines, whose formation requires temperatures much higher than the planetary temperatures, and where the optical density of the planetary atmosphere (and thus the effective planet size) is not expected to change across such lines.  At such wavelengths, it must be a closely valid approximation to treat the planet as an opaque body, whose size is indicated by transit photometry in adjacent passbands.

\subsection{Stellar properties}

The Alopex effective temperature is close to that of the well-studied K-giant Arcturus and their spectra are rather similar.  This resemblance to the extensively charted spectrum of Arcturus greatly facilitates the identification of spectral lines in Alopex, further eased by its low rotational velocity \citep{ceglaetal16, strachananglada17, torresetal12}, which produces relatively unblended lines despite a dense spectrum.

Overall stellar parameters have been deduced and compiled by numerous authors.  By combining interferometric observations of the limb-darkened angular diameter with several other measurements, \citet{boyajianetal15} deduced $T_{\textrm{eff}}$ = 4875\,$\pm$\,43 K, a stellar linear radius R$_{\star}$ = 0.805\,$\pm$\,0.016 R$_{\odot}$, and log~$\varg$ [cgs] = 4.56\,$\pm$\,0.03.  The metallicity is given by \citet{torresetal08} as [Fe/H] = --0.03\,$\pm$\,0.08.  

For the identification of many hundreds of spectral lines in Alopex, spectrum atlases of Arcturus are used as a guidance.  We recall the main parameters for that giant star as $T_{\textrm{eff}}$ = 4286\,$\pm$\,30 K and log~$\varg$ [cgs] = 1.66\,$\pm$\,0.05 \citep{ramirezallendeprieto11}.  Although this, of course, is a very different object, its temperature and spectrum are sufficiently close to those of Alopex, as to often enable straightforward line identifications, analogous to how lines in the G0\,V star HD\,209458 could be identified in Paper~II using solar spectrum atlases.

\subsection{M-dwarf companion}

Alopex is the primary of a double star system, accompanied by a mid-M dwarf, HD\,189733B, with a separation of 11.2 arcsec on the sky, corresponding to a projected distance of $\sim$216~au \citep{bakosetal06}.  Astrometric measurements suggest that the orbit of the secondary lies nearly in the plane of the sky (i.e., perpendicular to the orbital plane of the transiting planet) and has a period of $\sim$3200 years.  Such a wide binary does not perceptibly influence the surface structure of Alopex; also practically no spectrometer straylight is expected from this well-separated and faint companion, and there is no measurable change in stellar radial velocity during the hours of a planetary transit.

\subsection{Stellar activity}

Alopex is photometrically variable (classified as a BY~Dra type star) that has an active chromosphere \citep{boisseetal09, strassmeieretal00,winnetal07} and a magnetic field \citep{faresetal10}.   Long-term photometric monitoring shows somewhat erratic variations following a quasi-periodic cycle of around 12 days, with an amplitude of $\sim$3\,\% at visible wavelengths.  This is obviously the period of stellar rotation \citep{henrywinn08} and the amplitude of such light variations is equivalent to a Jupiter-sized dark spot with temperature $\sim$1000\,K cooler than the photosphere, rotating into and out of view.  The variability also manifests itself as a jitter in the apparent radial velocity on a level of $\sim$10 m\,s$^{-1}$ \citep{lanzaetal11, winnetal06}.  From space photometry, there is no evidence for any additional transiting planets, moons, nor rings \citep{crolletal07,pontetal07}. 

The presence of stellar activity offers both opportunities and problems.  Precision photometry during transit suggests that the planet sometimes crosses some dark spots \citep{pontetal07,singetal11}, while bright flares may appear in chromospheric emission lines \citep{czeslaetal15, klocovaetal17}.  A challenging future task will be to recover spatially resolved spectra of such starspots and active regions (including their magnetic signatures), but that will require simultaneous sequences of precision photometry and spectroscopy, which are not yet available.  For the present project, it introduces a slight uncertainty from using an average transit light curve, while the spectra originate from different transits.  Some recordings could possibly be from instances when the planet happened to cross some starspot, and the error budget for the retrieved spectra would then be enhanced by an unknown ratio of contributions from the magnetic and quiet photospheres.  However, Alopex is not strongly variable (especially not on short timescales comparable to the transit times) and its starspot area coverage is small, likely adding only a minor contribution to the error budget.   As seen in photometric light curves spanning several 12-day rotation periods \citep{strassmeieretal00}, the variability almost entirely follows the rotation period and has only small fluctuations on shorter timescales.  However, without simultaneous photometry, any recovery of spatially resolved chromospheric emission lines would be uncertain and will not be attempted here.

\section{Planet HD\,189733Ab}

The planet has been extensively studied with numerous instruments (not least in the infrared) for emission and transmission spectroscopy, inferring its atmospheric day-night thermal structure, a planetary hotspot near the substellar point, atmospheric winds, dust, molecules and atoms in the evaporating upper atmosphere, etc.  For this project, however, our main interest lies in the geometric size of the planet at those visual wavelengths at which our spectroscopic analysis is to be carried out, as well as knowing the transit path along the chord across the star.
Being slightly larger than Jupiter, the apparent radius of the planet was deduced from observations with the ACS instrument on the HST as 0.1572$\pm$0.0004~R$_{\star}$ \citep{pontetal07}, while various wavelength passbands in the HST STIS instrument yielded values between 0.15617 and 0.15754~R$_{\star}$  \citep{singetal11}.  Ground-based data show only very slight variation among various wavelength bands in the optical \citep{borsaetal16}.   Even in the far-infrared, the relative area covered by the planet (the square of these numbers, 2.46\,\%) remains about the same \citep{agoletal10,beaulieuetal08}.  For our later calculations, this implies that choosing the exact wavelength band for the photometric transit should not be critical.

\subsection{Planetary orbit and transit geometry}

The geometry of the planetary transit must be known to fully interpret any observed line-profile variations.  Through the Rossiter-McLaughlin effect, the position and path of the planet across the stellar disk can be determined.  Details of that method are explained in, for example, \citet{gaudiwinn07} and \citet{hiranoetal10, hiranoetal11}.  Different authors deduce slightly different values for the various parameters but in a critical review of compiled data, \citet{torresetal08} confirmed that the planet moves in an almost circular prograde orbit of period
2.21857312$\pm$0.00000076 days and has an inclination of 85.58$\pm$0.06$^{\circ}$.  That value is measured relative to the plane of the sky such that a value of 90$^{\circ}$ implies a transit across the center of the stellar disk.  The orbital radius of 8.863$\pm$0.020\,R$_{\star}$ gives an impact parameter (minimum sky-projected distance to stellar disk center, in units of R$_{\star}$) of $b$=0.6631$\pm$0.0023 \citep{agoletal10}. The planet reaches its point closest to stellar disk center at $\mu$ = cos$\theta$ = 0.75, where $\theta$ is the angle between the vertical from the stellar surface and the direction toward the observer.  It is somewhat more difficult to determine the inclination of the stellar rotation axis to the plane of the sky.   \citet{henrywinn08} found a lower limit of 54$^{\circ}$, which is consistent with the stellar spin being aligned with the planetary orbit.  More recent discussions of specific stellar parameters are by \citet{ceglaetal16,colliercameronetal10} and \citet{triaudetal09}.

\begin{figure}
\centering
\includegraphics[width=\hsize]{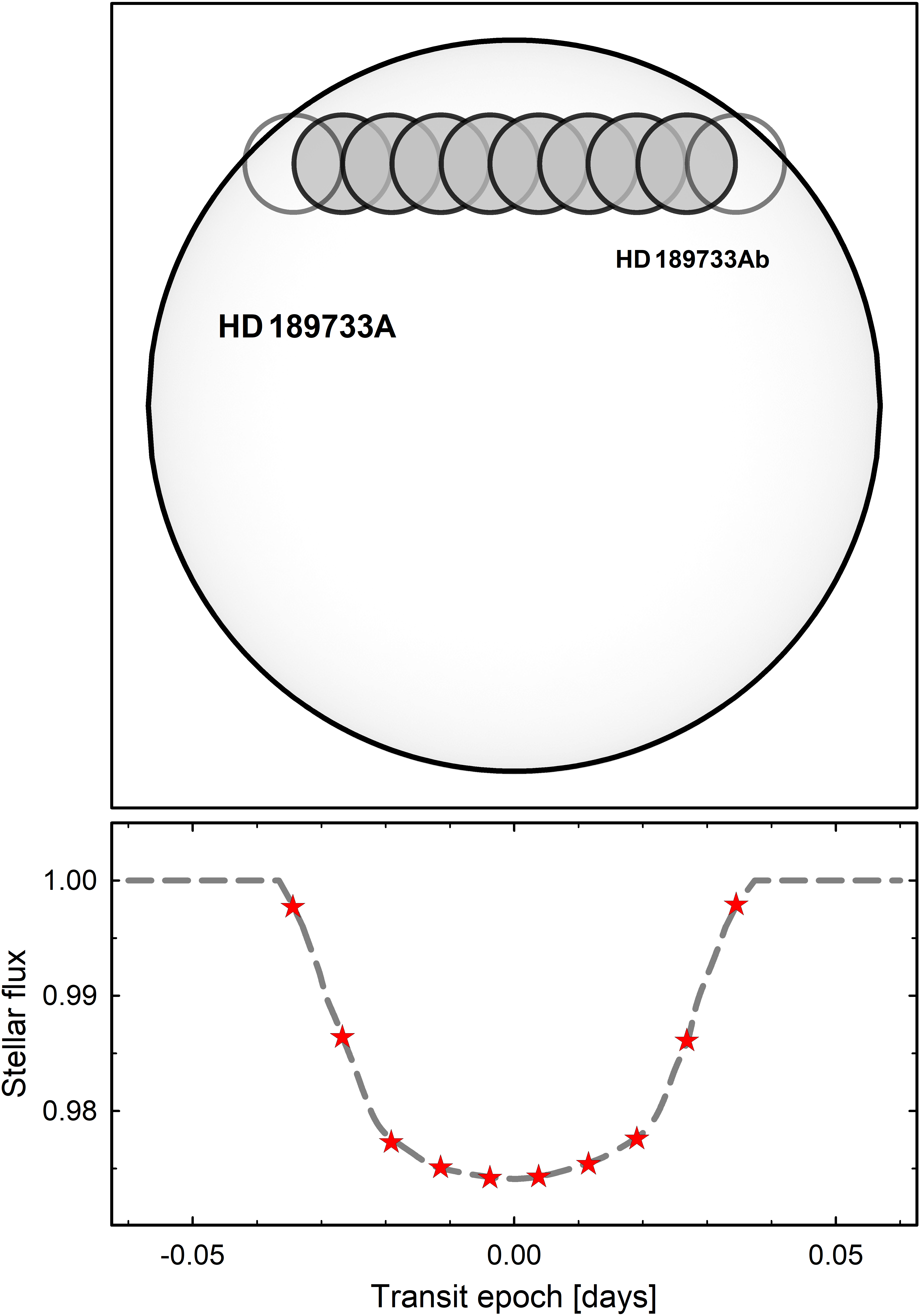}
\caption{Top: Exoplanet transit geometry.  Numerous spectral exposures were combined into ten average transit epochs, shown with the planet positions to scale (time increases left to right).  Observations close to the limbs did not reach sufficient S/N for the later spectral reconstruction, and those two epochs are shown as open circles.  Bottom: Phase-folded HST light curve combining three planetary transits by \citet{pontetal07} with the flux at each epoch indicated by stars. }
\label{fig:planet_positions}
\end{figure}

\subsection{Photometric transit}

The time of each mid-transit is obtained from the ephemeris as $T_i$ = $T_0\pm iP$, where $T_0$ = 2453988.80336 MJD is a reference mid-transit time,  P = 2.21857312~d is the orbital period of the planet \citep{extrasolarplanets17}, and $i$ = 1, 2, 3, ... the sequential transit event.  The maximum photometric transit depth reached during the 109.6-minute transit is 2.6\,\% in the visible.  With the impact parameter $b$ = 0.663, the transit occurs at stellar apparent latitude $34^{\circ}$ and the length of the transit path equals 0.749 stellar diameters.  With the planet diameter 0.157\,$\times$\,stellar, the transit proceeds with one planet diameter per $\sim$23 minutes, and thus exposures much shorter than that do not result in significant spatial smearing (except perhaps very close to the limb).

To retrieve the spectrum temporarily hidden behind the planet at each transit epoch, the fraction of the flux then removed by the planet must be known.  This fraction of hidden flux, or effective planetary area, $A_{\textrm{eff}}$ (in units of full-disk stellar flux), is the product of the local stellar brightness and the area of the planet.  Since the transits are repetitive, it is possible to use such data from different transit events from those of the spectroscopic measurements.  We obtained the values of $A_{\textrm{eff}}$ from the photometric light curve measured by \citet{pontetal07} with the ACS instrument on the HST (Fig.\ \ref{fig:planet_positions}).  Of course, this value could also be calculated from the transit geometry and limb-darkening law of the star; however our procedure is model independent and does not assume any stellar limb-darkening model.  As remarked above, because of its atmosphere, the effective size of the planet may slightly vary with wavelength.  However, the photometric transit observations with the HST ACS cover 550--1050 nm, largely overlapping our spectral region.  Actually, the measured transit light curves differ only slightly between different optical passbands, and reconstructed line profiles were found in Paper~II to be not sensitive to modest errors in effective planetary area.

\subsection{Barycentric motion and astrocentric correction}
 
The measured velocity semi-amplitude induced on the host star by the planet over its orbital period is K = 205\,$\pm$\,6 m\,s$^{-1}$ \citep{bouchyetal05}.  Since the eccentricity is close to zero \citep{boisseetal09}, i.e., the orbit is nearly circular, the barycentric radial-velocity motion of the star can be closely approximated by a sinusoid with amplitude 205 m\,s$^{-1}$.  Such a function was applied to adjust the heliocentric velocities to astrocentric values, thus placing the stellar spectra on a static frame-of-reference relative to the stellar center of mass.

\section{Spectroscopic data}

As detailed in Papers I and II, the signal-to-noise requirements limit usable data to the highest fidelity spectra from high-resolution spectrometers.   For example, if one desires a signal to noise in the reconstructed spectrum on the order of 100, say, extracted from only $\sim$2\% of the total stellar signal, this requires an original S/N around 5,000.  This is compounded by the center-to-limb decrease of stellar surface brightness.  While its amount depends on the wavelength region, it is more pronounced for cooler stars, there also enhanced by absorption-line contributions.  The limb darkening for Alopex was studied by \citet{hayeketal12}, who found that the surface brightness in the U, V, and I photometric bands decreased to 70, 80, and 87\% of their disk-center values at half a stellar radius from disk center, and to 35, 50, and 66\% at 0.9 radii, making near-limb measurements particularly challenging.  The observed signal was shown in Fig.\ \ref{fig:planet_positions} while modeled continuum values for other stellar models are in Fig.\ \ref{fig:3d_model_profiles} below.

Fortunately, Alopex is among the objects of which extensive observations have been carried out and numerous spectra are available in observatory archives.  For the study of stellar photospheres, spectra in the visual region are preferred: in the ultraviolet there is too much stellar line blending and in the infrared the paucity of stellar lines is compounded by telluric absorption bands. 

After searching the archives of several major observatories, suitable data sets covering the planetary transit were identified from the ESO UVES and HARPS spectrometers.  From these, two observing runs with HARPS (High Accuracy Radial velocity Planet Searcher) were selected.  Having two independent data sets enables both the combination of their data, and also the ability to check the results from one against the other.  The HARPS instrument is a high-resolution ($\lambda$/$\Delta\lambda$ $\approx$~115,000) spectrometer with an excellent wavelength stability on the order of 1~m\,s$^{-1}$  \citep{mayoretal03}.  Light from the ESO 3.6 m telescope is fiber-fed to an echelle spectrometer covering 380--690 nm over 72 echelle orders, recorded onto two CCD detectors.  The data selected originate from two observing runs on July 19 and August 28, 2007, respectively.  Each of these contain sequences of 20 exposures during the actual transit, each with 300~s exposure time, as well as a similar number of exposures outside transit.  During the hours of transit, as observed through airmasses around 1.7, photometric signal-to-noise ratios (as computed by the data reduction pipeline for the better-exposed echelle orders) were drifting around 100--120 per spectral resolution element.  For more details, see \citet{gustavsson17}.

\section{Data analysis}

Extracting a stellar spectrum from the small area covered by a transiting planet requires a S/N much higher than available in individual spectral exposures.  Since photospheric lines that share similar parameters are formed in a similar manner within the stellar atmosphere, numerous such lines may be averaged to obtain a low-noise representative profile.  Similar to Paper~II, we selected \ion{Fe}{I} lines since these are numerous and have been studied and modeled in many other contexts.  A difference from the G0~V star HD\,209458 in Paper~I, however, is the significantly more crowded spectrum of the cooler Alopex, which offers more numerous lines but also poses some challenges for the precise line extraction.

\begin{figure}
\centering
\includegraphics[width=\hsize]{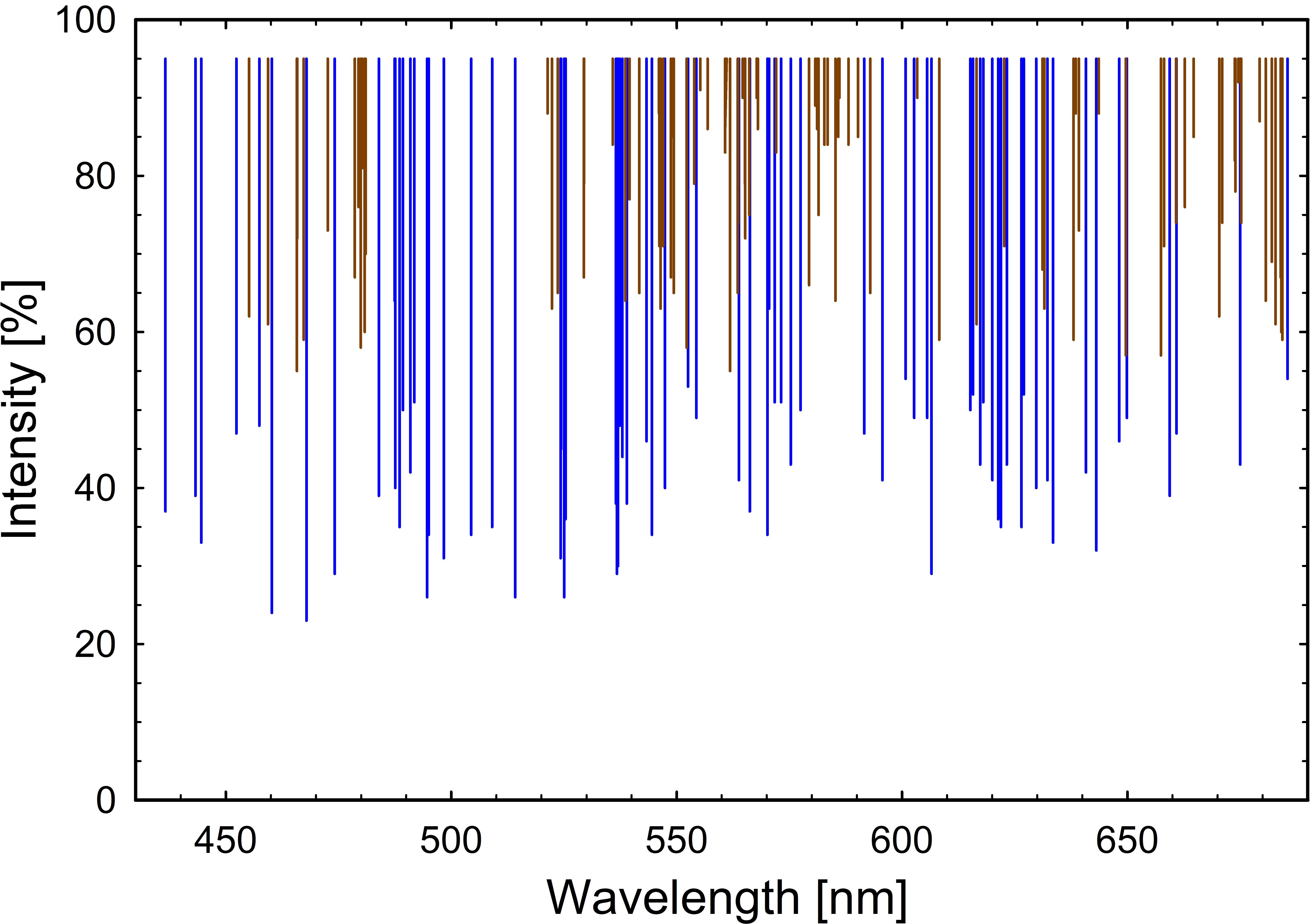}
\caption{Distribution of the 158 selected \ion{Fe}{I} lines.  Increased line blending below $\sim$450 nm makes locating unblended lines increasingly difficult.  Lines with measured residual central intensity value below and above 55\% are indicated in blue and red, respectively (forming strong and weak line groups).}
\label{fig:selected_lines}
\end{figure}

\subsection{Identifying spectral lines}

The temperature of the K1-dwarf  Alopex is not much different from that of the K0-giant Arcturus, one of the few stars with extensive line identifications in a high-resolution spectral atlas \citep{hinkleetal00}. Further, the metallicity of Alopex is not substantially different from that of the Sun, and they have many spectral lines in common.  Hundreds of lines need to be selected to reach the required S/N, and for their selection, we started with a listing of apparently unblended solar lines, verified their presence in the spectrum of Alopex, and also checked the identified lines against the Arcturus spectrum to remove those that can be suspected to be blended or contaminated by others.  Having a clear empirical footing, we prefer this procedure to, for example, using a synthetic spectrum from an atomic-line database run through a stellar atmosphere model.

\subsection{Selecting \ion{Fe}{I} lines}

We started with the list of 402 largely unblended solar \ion{Fe}{I} lines by \citet{stenflolindegren77} to identify prospective lines.  Out of these, 373 could be identified  in the spectrum of Alopex, while slightly less than half (158) were found to be reasonably unblended.  Their distribution is shown in Fig.~\ref{fig:selected_lines}, plotted as observed absorption down from the continuum.  Detailed line data are given in the Appendix Tables A.1 and A.2.

\subsection{Continuum determination}

\begin{figure}
\centering
\includegraphics[width=\hsize]{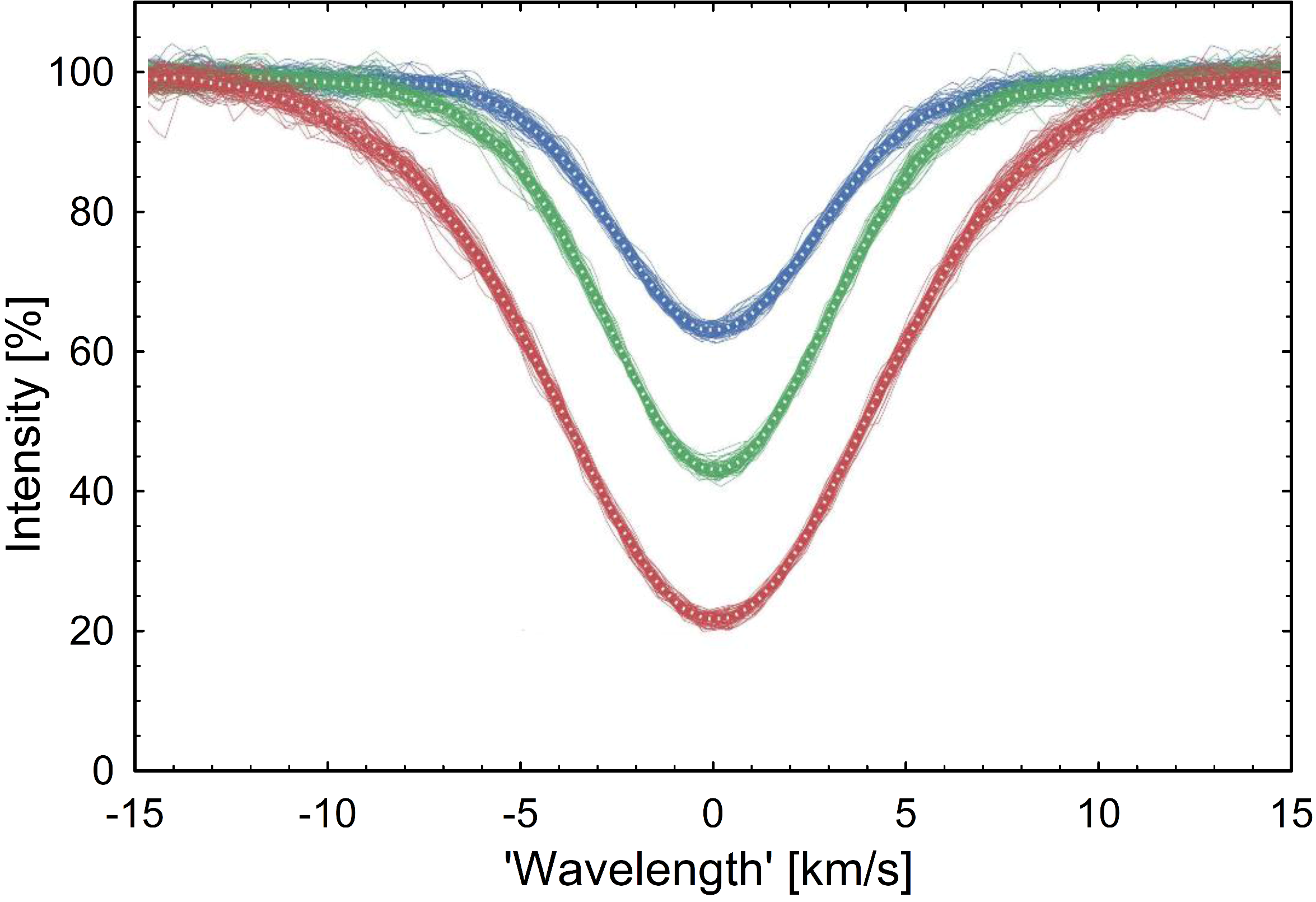}
\caption{Three representative \ion{Fe}{I} lines recorded in many successive exposures and their averages.  Individual profiles are shown as solid curves with their averages dotted.  From top down, the lines are \ion{Fe}{I} 680.685, 537.958, and 467.885 nm. }
\label{fig:averaged_lines}
\end{figure}

While the HARPS data do have exquisite wavelength stability, the photometric retrieval of continuum-normalized line profiles is somewhat complex because of the rapidly varying intensities in the dense spectrum where often no true stellar continuum is reached next to the line, which is why a wider interval has to be examined.  To limit the effort, a standardized procedure was applied.  After locating a spectral interval of $\pm$0.15 nm around any candidate spectral line, we found a first-order estimate of the local continuum by first normalizing to the largest local intensity value and then dividing by the median of flux values above 95\% in intensity.  For these selections, regions close to the strongest lines were avoided. To find the central wavelength of a line (in order to enable its later placement for averaging on a common velocity scale), we carried out a nonlinear least squares fit with a five-parameter Gaussian function to the immediate vicinity of the line (covering $\sim$0.04 nm), from which values for the local continuum and of the line center position were obtained.  This is of the type $I + (\lambda-\lambda_0)\Delta I/\Delta\lambda - I_0\cdot \exp\{-[(\lambda-\lambda_\star)/w]^2\}$, where $I$ is the continuum level, $\Delta I/\Delta\lambda$ the slope of the continuum, $\lambda_0$ the nominal wavelength from the line list, $I_0$ the line-center intensity, $\lambda_\star$ the line-center wavelength, and $w$ the line width.  Using these values, we next transformed the wavelength scales to equivalent Doppler velocities, applying the astrocentric wavelength correction as above, and placing all lines on a common intensity and wavelength scale.  Fig.\ \ref{fig:averaged_lines} shows a sample of the superposition of profiles of a few \ion{Fe}{I} lines from many successive exposures, illustrating typical levels of reproducibility and noise.

\subsection{Line averages and reference profiles}

\begin{figure}
\centering
\includegraphics[width=\hsize]{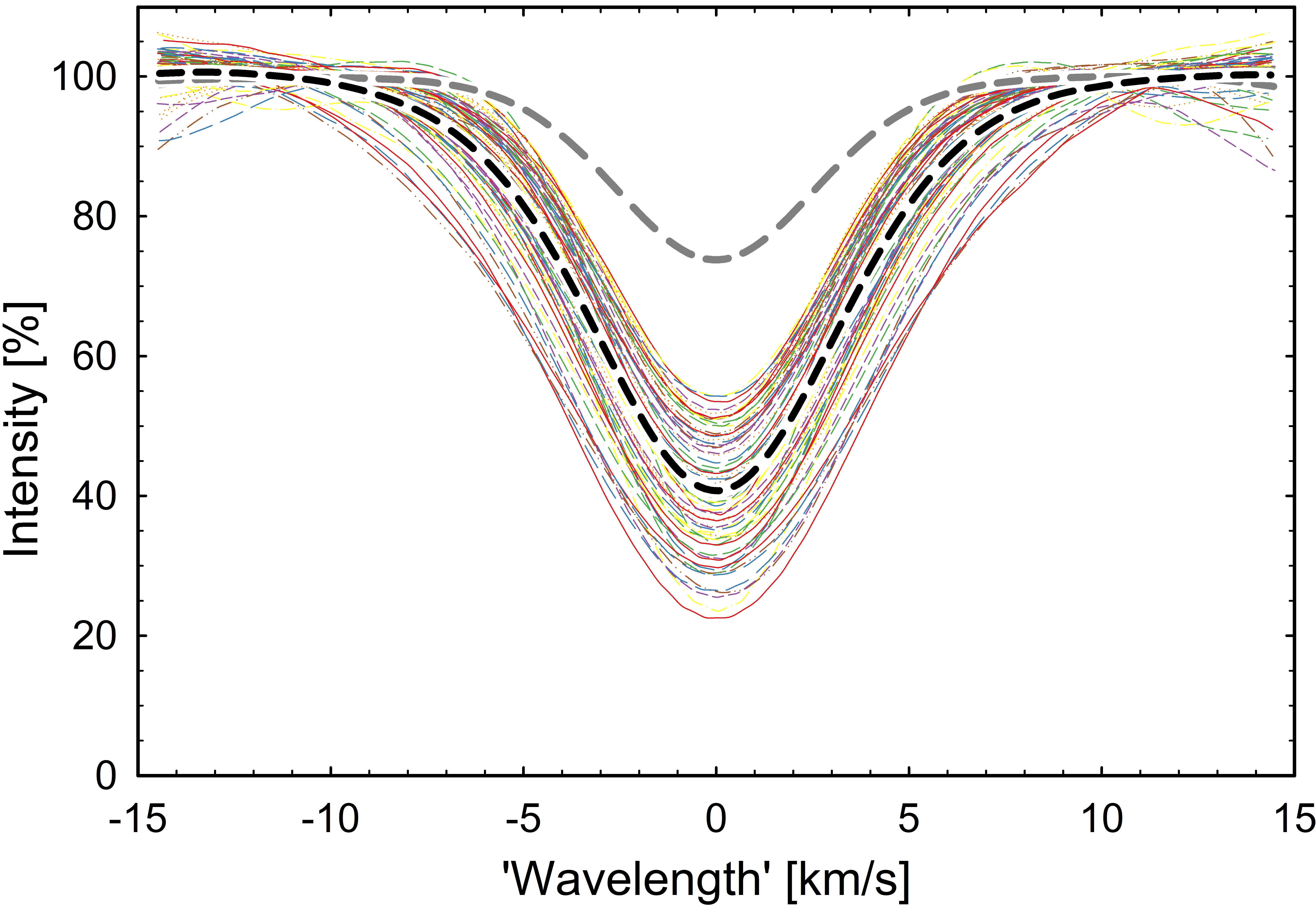}
\caption{Forming averages from groups of similar lines.  Individual profiles of the 72 selected strong \ion{Fe}{I} lines (observed central intensities below 55\% of the continuum) are shown (colored), together with their average (dashed black), forming a representative strong line profile.  An analogous subset of 86 weak \ion{Fe}{I} lines (central intensities shallower than 55\%) produces a representative weak line profile (dashed gray).}
\label{fig:all_lines_superposed}
\end{figure}

The group  of 158 selected lines was subdivided into two for strong and weak lines, respectively, splitting the groups at observed central absorption depths below or above 55\% of the apparent continuum.  By forming their averages (over 72 and 86 lines for the strong and weak groups, respectively), representative low-noise profiles for lines of different strength were obtained.  Figures \ref{fig:selected_lines} and \ref{fig:all_lines_superposed} show this line selection and their averaging.  For the strong-line group, the average wavelength is 558.3 nm, average line-depth intensity equals 40.9\% of the continuum, and average excitation potential $\chi$= 3.26 eV.   For the weak-line group, the averages are 581.6 nm, 73.6\%, and 3.77 eV.  The respective standard deviations in the wavelength distribution for the strong- and weak-line groups are 65 and 66 nm, and 1.07 and 0.89 eV for the excitation potentials (Tables A.1. and A.2.).  We note that these are the observed line strengths, in which the intrinsic stellar spectrum has been degraded by the finite spectrometer resolution.   

The averaging of line profiles is carried out in velocity space, individually scaling each observed line profile.  After finding the position of any line center, relative wavelength positions are converted to equivalent Doppler velocities, and the intensities linearly averaged.  We note that this would be equivalent to averaging in Fourier space as applied by, for example, \citet{gray16}, only with possible slight differences in how noise properties are propagated. 

Experiments were also made with various subsets of these lines, all of which must of course be blended to at least some small extent.  For example, the same analysis as  below was carried out by retaining only smaller subsets of more strictly unblended lines.  If required to be surrounded by an apparent continuum of at least 95\% intensity over $\pm$15 km\,s$^{-1}$, 74 out of these 158 lines remained.  However, in the end, no significant difference in results was seen, suggesting that the weak blends present seem to be of random character and do not contribute systematic effects.

The averages of lines in Fig.\ \ref{fig:all_lines_superposed}, with each individual profile starting with S/N~$\gtrsim$100, begin to approach signal-to-noise ratios on the order of 1,000 for each exposure (i.e., for the stochastic part of the noise).  However, this is still not adequate and additional averaging is needed to further enhance the signal.  Half of the about 80 exposures from the two observing runs are recorded outside transit, and these are next averaged to produce a low-noise reference spectrum against which the spectra recorded during transit are to be contrasted. 

By first forming such an average over all 20+19=39 exposures recorded outside the two transits, and then over the many lines in each line group, averaging is achieved over $\sim$2,500 profiles and the nominal S/N can be expected to reach $\sim$5,000.  Equally low-noise spectra cannot be obtained for each transit phase, however.  With 20 in-transit exposures for each of the two observing runs, a total of 40 exposures are available over the transit chord.  To push the S/N, we divide the transit chord into 10 epochs and, for each of these, collate two exposures from each observing run to obtain the average of four exposures from each transit epoch; this increases the critical S/N by a factor two at the cost of somewhat decreasing the spatial resolution.

\subsection{Line profile ratios during transit}

\begin{figure}
\centering
\includegraphics[width=\hsize]{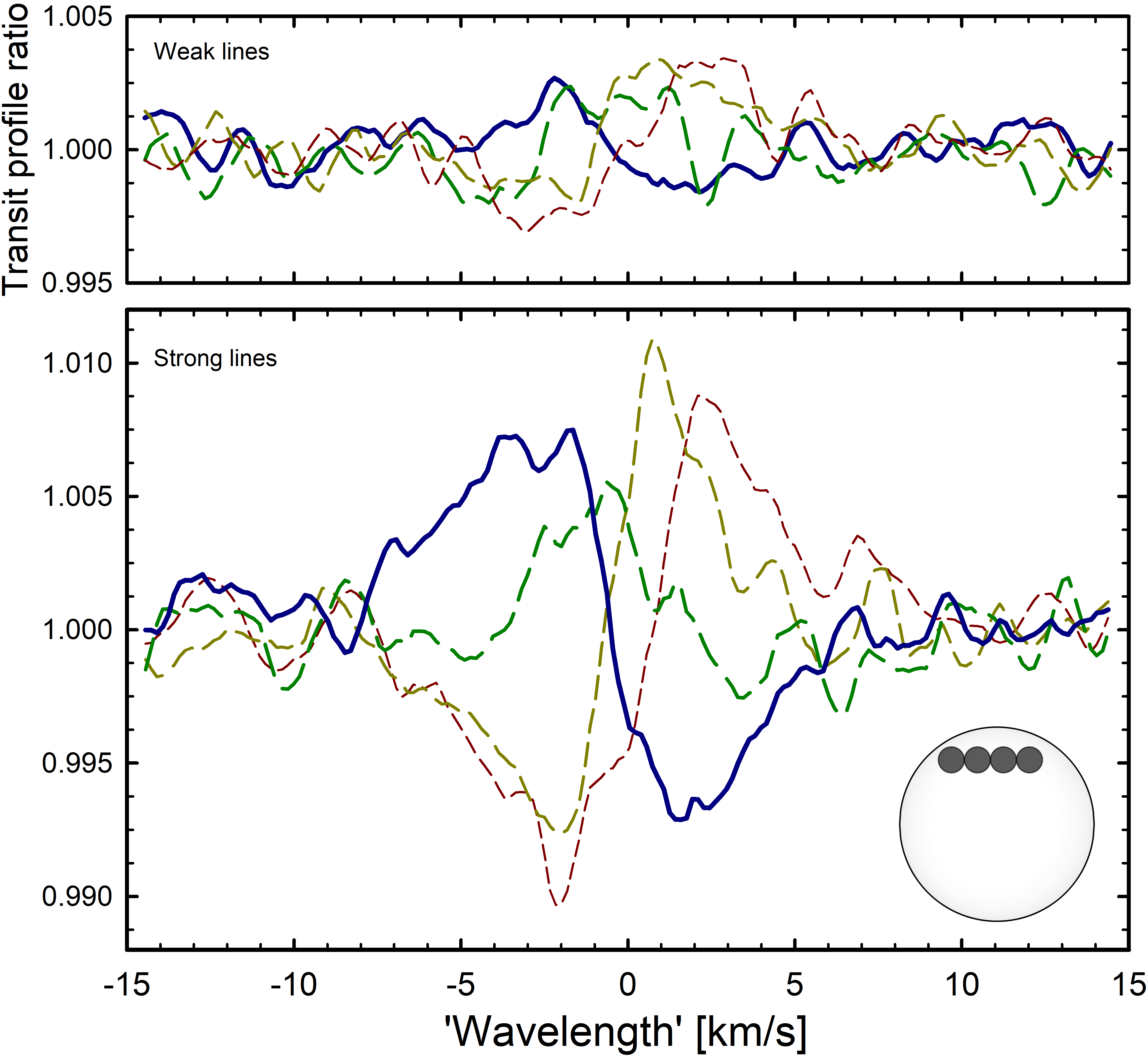}
\caption{Ratios between observed weak and strong line profiles to the profile from outside transit, at four different transit epochs, indicated by planet positions to scale across the stellar disk (time increases left to right).  Data for successively later times are plotted with successively shorter dashes. }
\label{fig:transit_line_ratios}
\end{figure}

The change in the spectrum during the various transit epochs relative to the full-disk spectrum outside transit is tiny and not easily discernible in an ordinary line-profile format. Fig.\ \ref{fig:transit_line_ratios} instead shows the ratios of line profile intensities at each different transit epoch relative to the reference profile from outside transit (for clarity, only every second transit epoch is shown).  This is the same format as used for theoretical predictions in Paper~I and for the observations of HD\,209458 in Paper~II.  For the strong lines, these ratios reach a maximum amplitude of around 1\,\% but rather less for the weak lines.  Although qualitatively similar, the quantitative amplitude depends on the intensity gradients in the line flanks: shallower profiles with weaker gradients produce smaller ratios, thus requiring better S/N for full analysis. A number of such synthetic line ratios for different models were in Paper~I.  In the ratios between our in-transit spectra with S/N $\sim$2,000 and the out-of-transit reference spectrum with S/N $\sim$5,000, noise levels on the order of 0.001 or slightly better are expected, consistent with the wiggles seen in Fig.\ \ref{fig:transit_line_ratios}. 

\subsection{Rossiter-McLaughlin effect}

\begin{figure}
\centering
\includegraphics[width=\hsize]{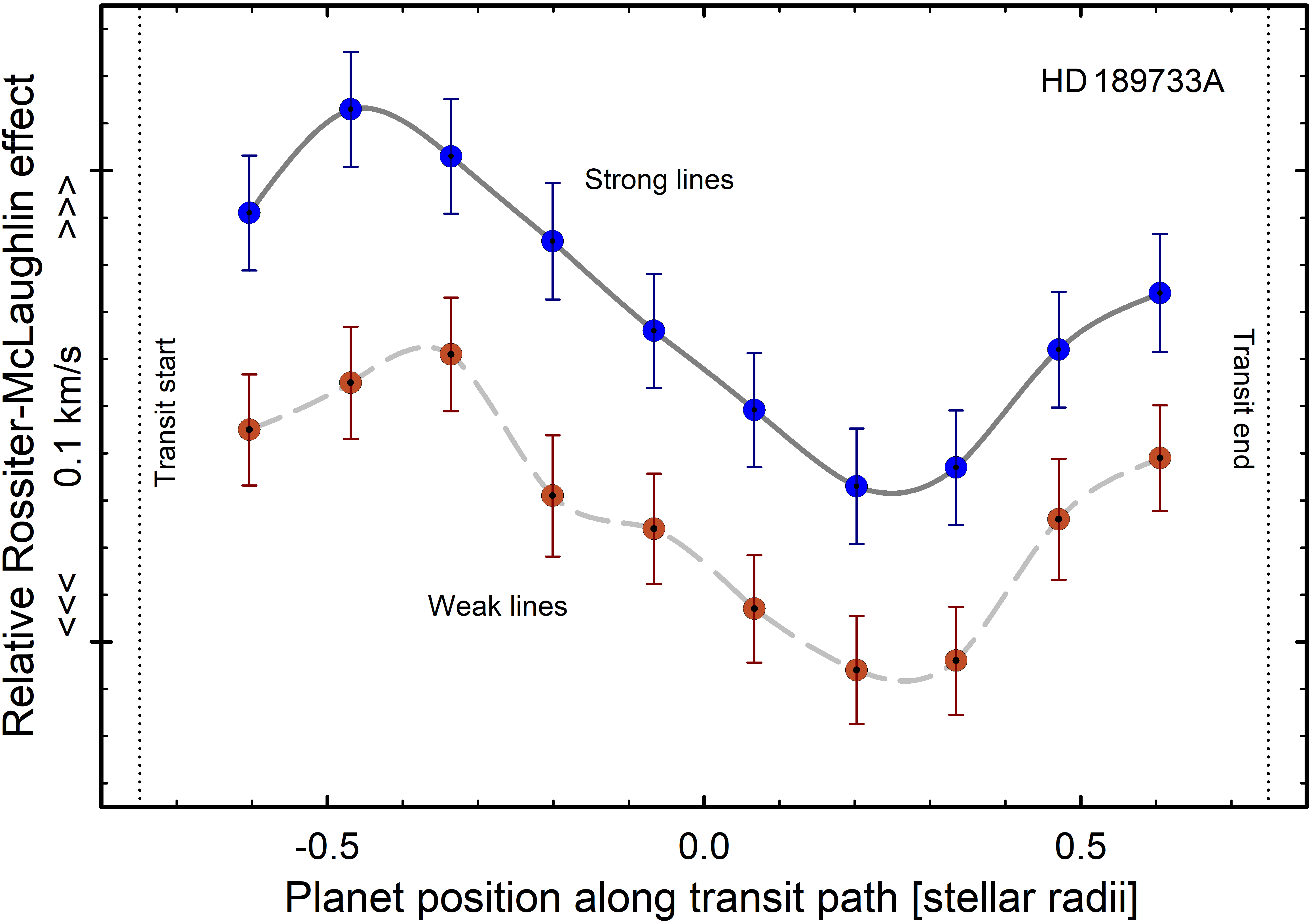}
\caption{Rossiter-McLaughlin effect of the apparently changing stellar radial velocity during transit is recovered from fitting line-center wavelengths to the strong and weak \ion{Fe}{I} line averages.  Velocity increases upward but shifts are only relative.}
\label{fig:rossiter-mclaughlin}
\end{figure}

The overall sinusoidal patterns in Fig.\ \ref{fig:transit_line_ratios} originate mainly from the successively shifting line wavelengths during transit.  This modulation of the apparent radial velocity constitutes the Rossiter-McLaughlin effect, usually determined by cross-correlating the full stellar spectrum but here we can extract it for specific spectral lines from wavelength fits during the various transit epochs (Fig.\ \ref{fig:rossiter-mclaughlin}).  Although such line-center fits to \ion{Fe}{I} lines need not give identical values to those obtained from cross-correlations over entire spectra, the similarity to the measured Rossiter-McLaughlin effect in Alopex \citep[e.g.,][]{winnetal06} serves to confirm the integrity of the present data reductions. However, these measured values only show relative velocities.  As illustrated by various theoretically modeled Rossiter-McLaughlin curves in Paper~I, a lot of additional information could be extracted if such curves could be obtained on an absolute wavelength scale and also separately for different types of spectral lines.  That endeavor will require very accurate laboratory wavelengths for many lines but appears feasible to attempt with data not much superior to the present set.

In a related but different approach, \citet{colliercameronetal10} used observations of Alopex to model the velocity cross-correlation function of the stellar spectrum to identify spectral-line distortions across the surface of the star during transit, deducing the trajectory of the missing starlight (`Doppler shadow') and obtaining measures of the projected stellar rotation rate and geometry.

\section{Spatially resolved line profiles}

The normalized spectral flux $S$ behind the effective planetary area $A_{\textrm{eff}}$ is obtained as $(S_{\textrm{outside-transit}}$ -- $S_{\textrm{in-transit}}$)/$A_{\textrm{eff}}$, while the steps in reconstructing continuum-normalized spatially resolved profiles were described in Paper II (e.g., its Fig.\,9).  Such profiles obtained from the strong- and weak-line averages are shown reconstructed at various stellar disk positions in Figs.\ \ref{fig:reconstructed_strong} and \ref{fig:reconstructed_weak}.  
These are for the transit epochs 2--9, where the nominal S/N in the reconstructed profile intensities near mid-transit is around 35, falling to $\sim$20 at the beginning and end. 

Since the reconstructions involve dividing subtracted spectra (with noise $\lesssim$10$^{-3}$ of the continuum) by the small values of $A_{\textrm{eff}}$ (Fig.\ \ref{fig:planet_positions}), the noise is correspondingly enhanced.  Closer to disk center, $A_{\textrm{eff}}$ reaches values of $\sim$0.025, enabling S/N ratios on the order of 30.  However, at the near-limb positions at epochs 1 and 10, $A_{\textrm{eff}}$ drops to very small values.  Those profiles attain much lower S/N and are not shown.  Measuring near-limb positions is hard both because the star is substantially limb darkened and the planet is already partially outside the stellar disk.

\subsection{Stellar rotation}

\begin{figure}[H]
\centering
\includegraphics[width=\hsize]{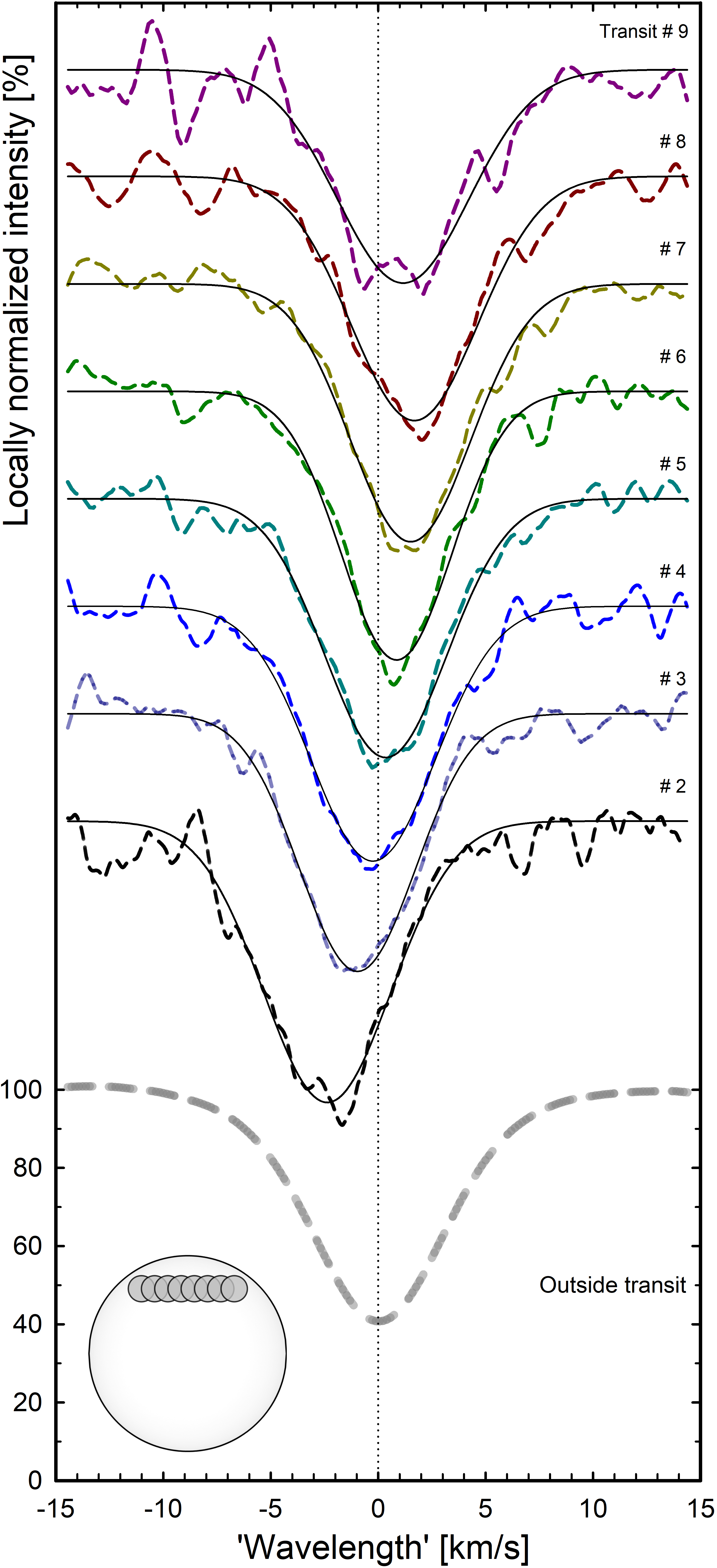}
\caption{Reconstructed line profiles (bold dashed) for the strong \ion{Fe}{I} line at eight positions during the planetary transit, and Gaussian fits (thin solid). Spatially resolved profiles are not subject to rotational broadening and are deeper and narrower than the disk-integrated profile outside transit shown at bottom.  The successive transition from blue- to redshift shows the prograde orbital motion of the planet relative to stellar rotation.  }
\label{fig:reconstructed_strong}
\end{figure}

\begin{figure}[H]
\centering
\includegraphics[width=\hsize]{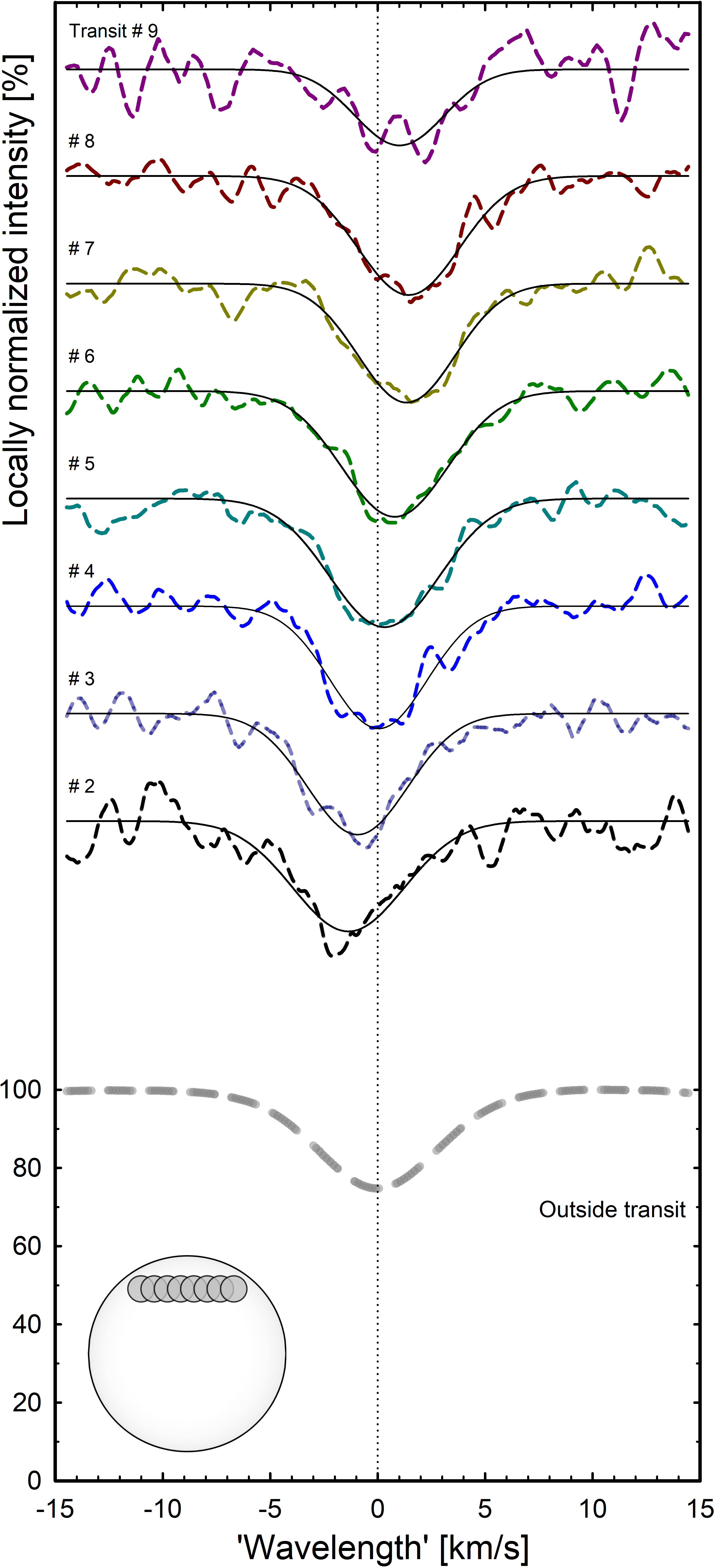}
\caption{Reconstructed line profiles for the weak \ion{Fe}{I} line at eight positions during the planetary transit, analogous to Fig.\ \ref{fig:reconstructed_strong}.  Time increases from bottom up, and the planet positions for the successive transit epochs (left to right) corresponding to each line profile are to scale.}
\label{fig:reconstructed_weak}
\end{figure}

\begin{figure*}
%\sidecaption
 \centering
 \includegraphics[width=18cm]{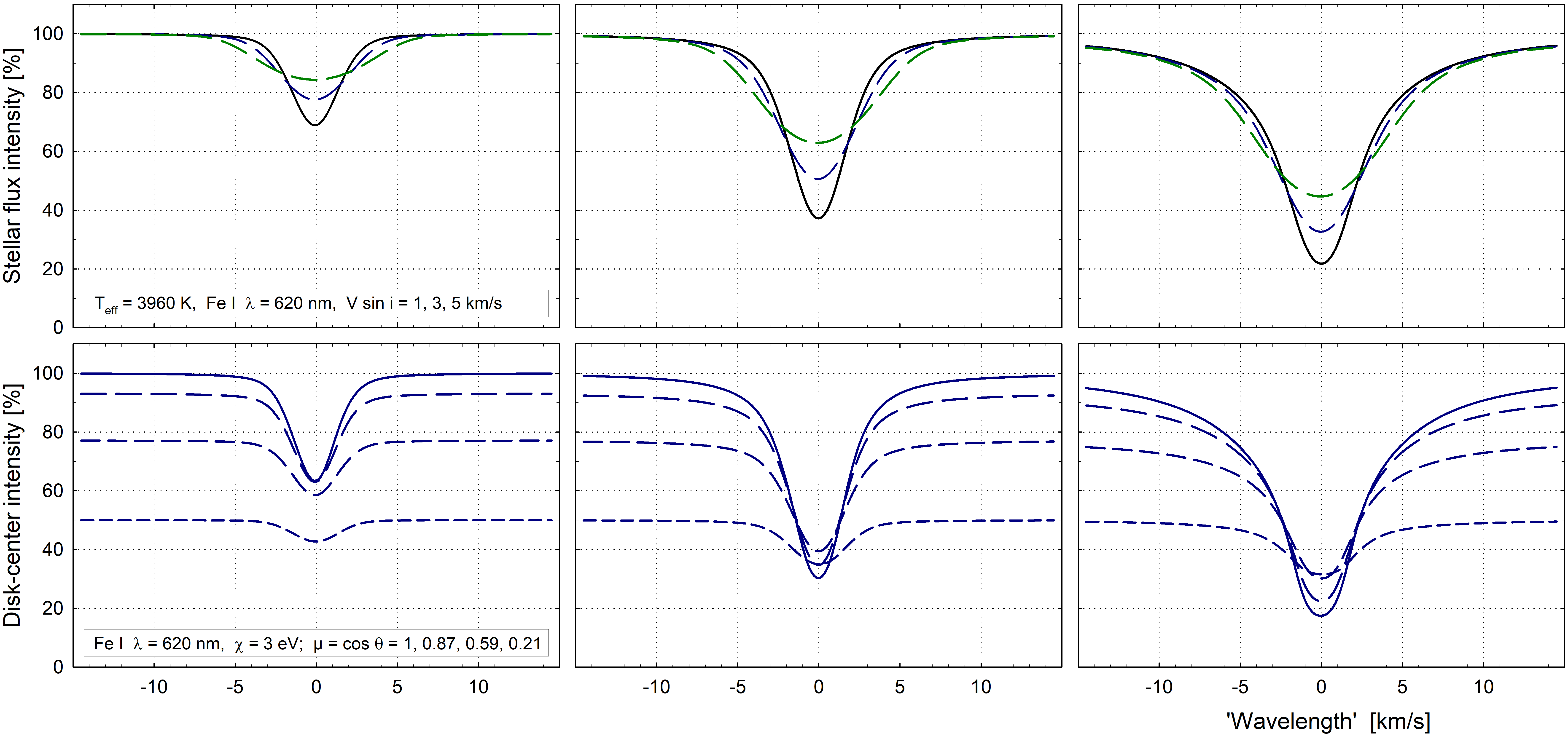}
     \caption{Synthetic \ion{Fe}{I} profiles for lines of different strengths, computed from a CO\,$^5$BOLD model of a cool K-type dwarf with $T_{\textrm{eff}}$ =3960~K.  Top row shows full-disk profiles for stellar rotational velocities V\,sin$i$= 1, 3 and 5 km\,s$^{-1}$; bottom row represents center-to-limb changes for these profiles.  } 
     \label{fig:cool_dwarf_profiles}
\end{figure*}

Not being subject to stellar rotational broadening, spatially resolved line profiles are deeper and narrower than the integrated profile from the full disk. The stellar rotation is known previously from both measured line broadening and from the Rossiter-McLaughlin effect but can be verified anew.  The gradual shift of the central wavelength shows the amount of stellar rotation at the latitude of the transit chord, while the shift from a relative blueshift on the ingress side of the transit, to a redshift on the egress side confirms the prograde orbital motion of the planet relative to stellar rotation (Figs.\ref{fig:reconstructed_strong}--\ref{fig:alopex_results}).

The apparent latitude of transit is known to be $\sim$$34^{\circ}$; the tilt of the stellar rotational axis seems consistent with the stellar spin being aligned with the planetary orbit \citep{henrywinn08}.  The velocity shift of $\sim$3 km\,s$^{-1}$ observed across the (partially measured) planetary transit chord at a latitude where the projected stellar rotational velocity equals cos\,$34^{\circ}\!\!\approx$\,0.8 times the equatorial value, is consistent with various previous determinations indicating a comparable value for $V$\,sin$i$ but would need improved data with higher spectral resolution to be better constrained. 

\subsection{Fitted line parameters}

In addition to the reconstructed line profiles at different epochs, Figs.\ \ref{fig:reconstructed_strong} and \ref{fig:reconstructed_weak} also show Gaussian fits to these profiles.  By averaging over the remaining noise, such fits provide parameters of the overall line profiles, which are shown in Fig.\ \ref{fig:alopex_results}.   Line-bottom intensities are in units of the local spectral continuum at each $\mu$, while the line widths are obtained from five-parameter fits to modified Gaussian functions of the type $y_0 + a \cdot \exp[-0.5\cdot(|x-x_0|/W)^c]$, yielding $W$ as a line-width measure.  It is not suggested that the profiles would be intrinsically Gaussian, but such were found to give better fits than, for example, Lorentzian profiles.  These changes such as in fitted line-widths and line-bottom intensities can then be compared to theoretical expectations.

Figure \ref{fig:cool_dwarf_profiles} shows synthetic line profiles from a 3-D model of a K-dwarf photosphere (for modeling details, see Paper~I). Limb darkening is pronounced in these cooler stars, adding to the challenges of reconstructing line profiles close to the limb.  It can be noted that the strongest among these synthetic lines develops pronounced damping wings, while the shapes of the weaker lines are more Gaussian-like.  Our observational samples do not include any really strong lines.  In the initial scrutiny of the spectrum, a few such lines were identified as candidates but were not retained because their profiles were clearly outside the distribution in Fig.\ \ref{fig:all_lines_superposed}, while their few numbers would not permit us to reach meaningful S/N ratios. 

One center-to-limb signature that may be expected from the structure of solar-like granular convection comes from horizontal velocities in the photosphere generally being greater than vertical velocities.  Rising gases from below decelerate and turn over on heights comparable to the vertical pressure scale height, which is rather smaller than the horizontal extent of granules, and thus horizontal velocities must be greater to satisfy mass conservation.  Line widths would then show a characteristic increase toward the limb, where the greater horizontal motions contribute more Doppler broadening.  Such a gradual broadening is well known for the Sun \citep[e.g.,][]{lindetal17}, and was also indicated by the spatially resolved spectra across the G0~V star HD\,209458 in Paper~II.  However, no such effect is seen in the results of Fig.\ \ref{fig:alopex_results} (at least within the range of observed $\mu$).

\subsection{Spectrometer resolution}

Limitations in how rigorous analyses that can be carried out are set by not only the residual noise levels, but also by the finite spectral resolution and the imperfectly known spectrometer response function.  The exact instrumental profile of HARPS and its ensuing spectral resolution depend somewhat on the exact arrangement of the fiber that is feeding light into the spectrometer but is nominally $\lambda/\Delta \lambda\sim$115,000, i.e., $\sim$2.6~km\,s$^{-1}$ \citep{mayoretal03, rupprechtetal04, esoharps18}.  This value refers to the full width at half maximum, which in the spectrometer is sampled by about four detector pixels.  

When comparing to synthetic stellar profiles that are computed with much higher resolution, an appropriate correction has to be applied.  This correction implies a decrease of the observed line-width measures in Fig.\ \ref{fig:alopex_results} by $\sim$1.0~km\,s$^{-1}$ for the strong lines, and $\sim$1.2~km\,s$^{-1}$ for the weak ones.  Even then, the line widths would be slightly broader than their intrinsic values since another broadening arises from the spatial smearing across the stellar disk introduced by the averaging of a few sequential exposures.

\section{Theoretical 3-D simulations}

With currently achievable precision, comparisons to synthetic line profiles mainly have to be against their overall properties such as line widths or depths, rather than more subtle asymmetries or the dependence on excitation potential.  To understand what signatures that may be realistically detected by studies such as this work, such integral properties were examined for synthetic line profiles of main-sequence stars representing spectral types F, G, and K.  These are from our grid of CO\,$^5$BOLD model atmospheres \citep{freytagetal12} in the temperature range between $T_{\textrm{eff}}$ = 3960 and 6730~K, as described in Paper~I.  

No model in this grid has parameters precisely corresponding to those of Alopex, but it is bracketed by the models for 5900~K ($\sim$G0~V) and for 3960~K ($\sim$K8~V), both with solar metallicity and surface gravity log~$\varg$ [cgs] = 4.5.  Some earlier modeling of line profiles from 3\mbox{-}D models with parameters close to Alopex were made by \citet{dravinsnordlund90b} and \citet{ramirezetal09}, but those did not then foresee comparisons to spatially resolved spectroscopy.

\begin{figure}[H]
\centering
\includegraphics[width=8cm]{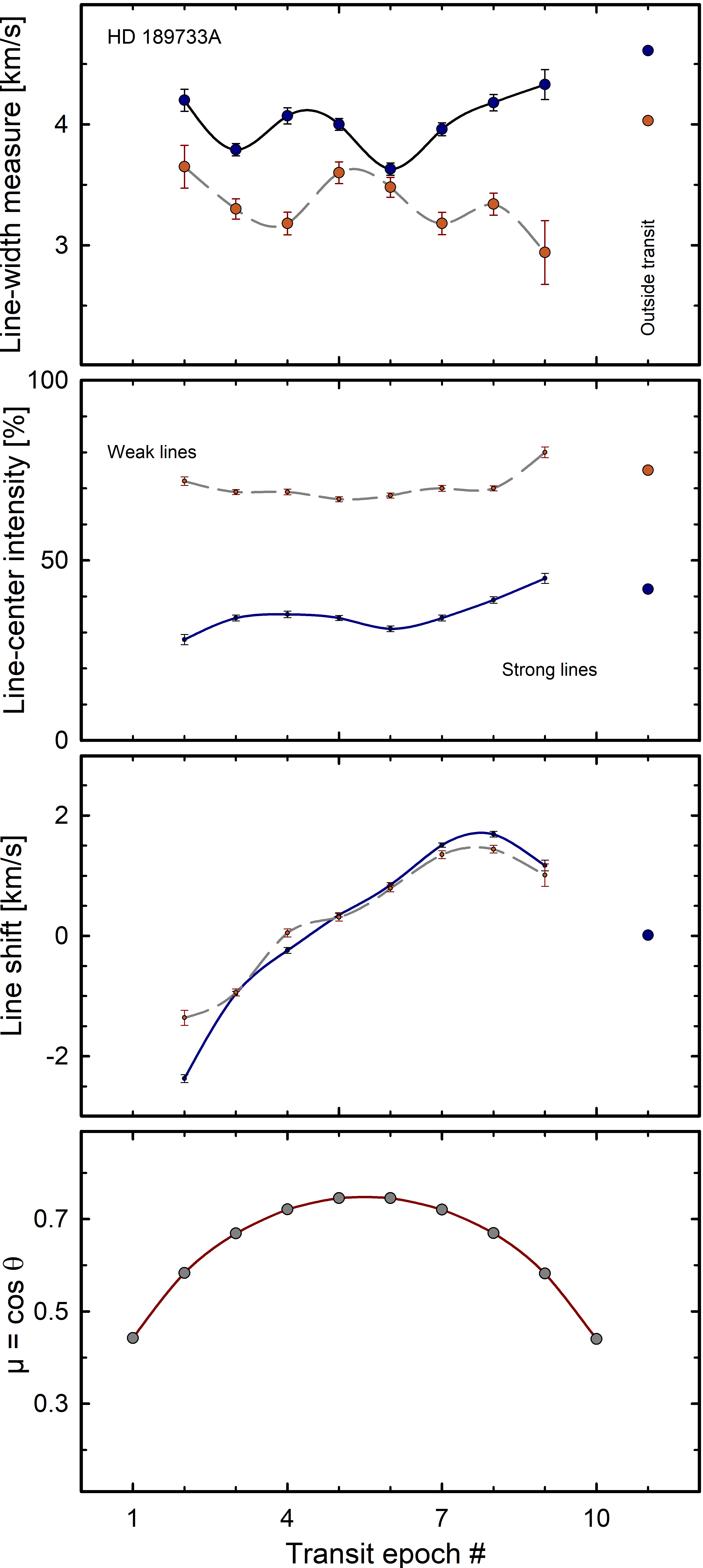}
\caption{Line-widths, line-bottom intensities, and wavelength shifts of reconstructed profiles for the strong (solid curves) and weak lines (dashed) at eight successive positions during the planetary transit, obtained from Gaussian fits to the data.  The error bars denote one standard deviation to these fits but judgments of any apparent dependences should also examine the reconstructed profiles to which these fits were made.  The various transit epochs, with their center-to-limb locations shown at bottom, correspond to those in Fig.\ \ref{fig:planet_positions}.  }
\label{fig:alopex_results}
\end{figure}

\begin{figure*}
%\sidecaption
 \centering
 \includegraphics[width=18cm]{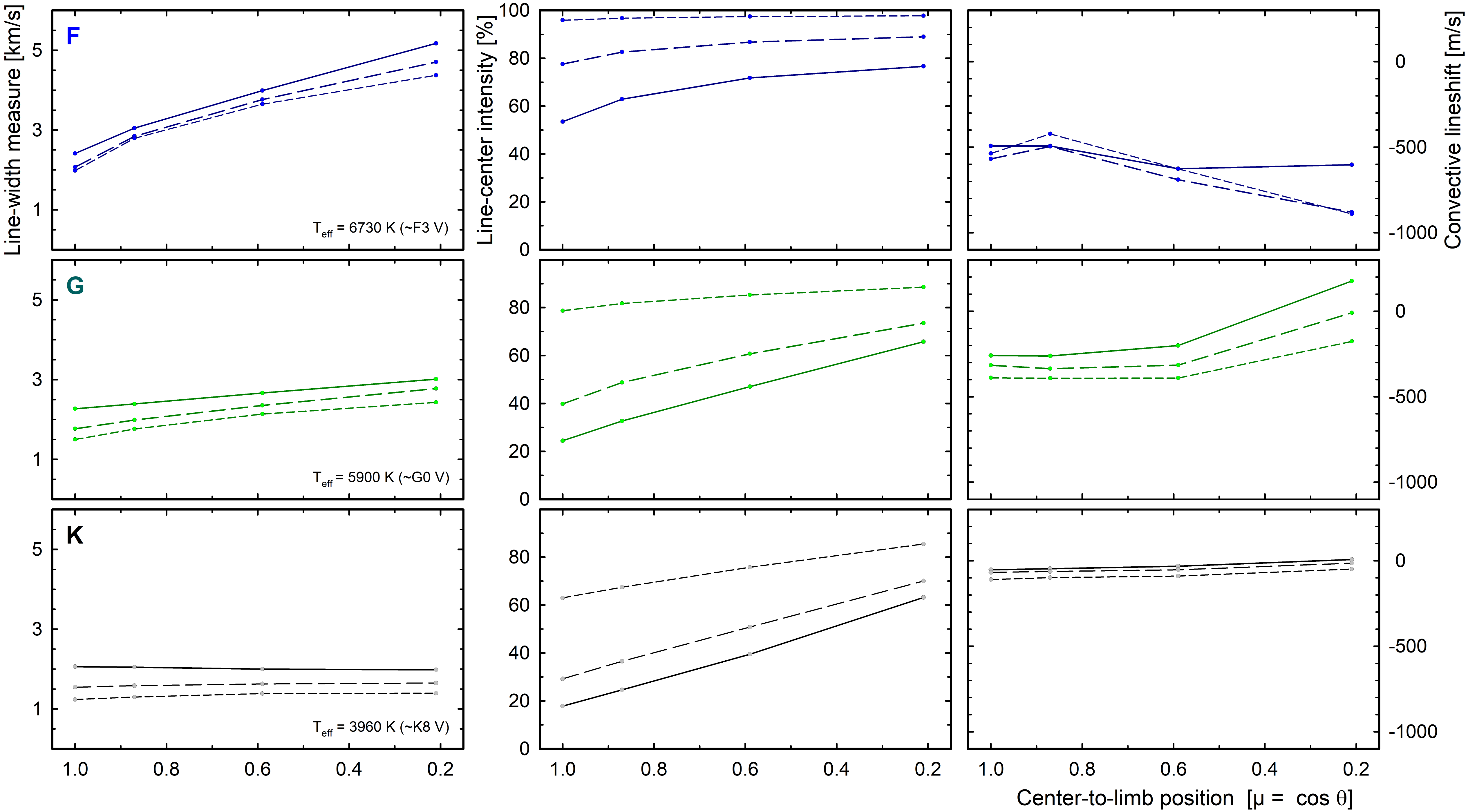}
     \caption{Center-to-limb changes of overall parameters for differently strong lines computed from the various CO\,$^5$BOLD models representing stars of spectral types F, G, and K.  Properties of strong lines are shown by solid curves, of intermediate lines by long-dashed, and of weak lines by short-dashed curves.  Line-width measures, line-center intensities, and convective line shifts are from symmetric Gaussian fits to the full synthetic profiles and therefore do not resolve any line asymmetries. Convective line shifts are in absolute units but gravitational redshifts are neglected. }
     \label{fig:model_center_to_limb}
\end{figure*}

In Fig.\ \ref{fig:model_center_to_limb}, the center-to-limb behavior of line widths, strengths, and shifts are shown for differently strong lines from various models.  As before, line-width measures are obtained from five-parameter fits to modified Gaussians while line-bottom intensities are in units of the local (limb-darkened) spectral continuum at each $\mu$.  Line-shift values obtained from symmetric Gaussian fits do not resolve asymmetric signatures such as bisector shapes, i.e., the varying line shifts at different depths in the line.  The convective line shifts are absolute (i.e., relative to the stellar frame of rest) but gravitational redshifts are neglected.  Similar to Paper~I, calculations are made assuming local thermodynamic equilibrium for \ion{Fe}{I} lines with wavelength $\lambda$=620~nm and lower excitation potential $\chi$=3~eV.  

In general, line widths are expected to increase toward the limb, where the greater horizontal motions should contribute more Doppler broadening.  However, the different surface structure of granulation in hotter and cooler stars strongly modifies the center-to-limb changes in line parameters, in particular those of the line width.  The vigorous granulation in hotter F-type stars is predicted to be naked, i.e., with full temperature and velocity contrasts directly visible across their corrugated optical-depth surfaces \citep{allendeetal02, chiavassaetal18, dravinsnordlund90a, dravinsnordlund90b}.  For example, the convective blueshift is predicted to increase from disk center toward the limb, where many spectral contributions come from horizontal flows on the sloping hills facing the observer.  These approaching flows are blueshifted while the equivalent redshifted components remain invisible behind these hills and an enhanced blueshift results.  This also exposes the full amplitude of the horizontal velocity field to an outside observer, causing a correspondingly greater line broadening. 

On solar-like G-type stars, the brightness and velocity amplitudes of granulation are slightly veiled beneath the line-forming layers.  As a consequence, the brightness contrast decreases with height, manifest in effects such as a clear dependence of convective blueshift on line strength and its decrease from disk center toward the limb.  In cooler K-type stars such as Alopex, however, the expected contrast of granulation that is visible on the surface is substantially lower than for the Sun or hotter stars since the granulation structures are more strongly veiled beneath the visible surface.  The fully developed contrast of granulation is not strictly a surface phenomenon in this case, but rather relates to those slightly deeper layers in which the energy flux changes from being mainly convective to mainly radiative.  Although this hidden granulation is not deep down in terms of linear depth -- maybe only some 100 km -- the corresponding optical depths are substantial for spectral line formation, and emerging line profiles are expected to carry but modest signatures of the convection in deeper layers \citep{chiavassaetal18, dravinsnordlund90b, magicetal13, nordlunddravins90, ramirezetal09, tremblayetal13}.  As a consequence, cool K stars are expected to produce only minimal center-to-limb changes in line widths (Fig.\ \ref{fig:model_center_to_limb}), which is consistent with the lack of any corresponding observed signature in Fig.\ \ref{fig:alopex_results}.  

With spectral lines only slightly affected by granulation, their variability due to changes in granulation must be expected to be small as well, and the ensuing tiny photometric and spectral microvariability appear to make such stars prime candidates in searches for Earth-analog exoplanets, for which the stellar microvariability is a limiting factor in hotter stellar hosts (at least as long as effects from magnetic activity, common in the coolest stars, do not become dominant).

\section{Outlook and future potential}

This work now extends the number of stars with high-resolution photospheric spectra across their surfaces to a total of three (including the Sun), and the experience gained in understanding what noise levels can realistically be reached, may guide the planning of future observations.  This work used different types of spectroscopic data compared to the study of HD\,209458 in Paper~II.  The wavelength stability of current HARPS spectra removed the need for such elaborate wavelength calibrations that were required for UVES spectra in Paper~II.  Current data have a slightly higher spectral resolution than those in Paper~II ($\sim$115,000 vs.\ 80,000) but not as good S/N in individual exposures ($\sim$100 versus\ 500).  However, the number of exposures during transit is greater (40 versus\ 14), as is the number of measurable lines (158 versus\ 37).  Also, the greater area of the planet (2.5\,\%  versus\ 1.5\,\%) contributes less noise in the spectral reconstruction, therefore the end results are comparable.  Still, we note that the range in $\mu$ for which line profiles could be reconstructed is lesser here than in Paper~II ($\mu$=0.75--0.58 vs.\ 0.86--0.24), both because of the transit geometry with a larger impact parameter, and a higher noise level near the epochs of transit ingress and egress, which constrains more detailed model fitting.

\subsection{Limitations in spectral-line averaging}

Present types of studies may ultimately be constrained by the number and types of spectral lines that actually are present in stellar spectra.  Even if theoretical models permit the precise calculation for lines of some particular strength, excitation potential, and ionization level from various atomic species, comparisons to observations may be unfeasible if real stars would not possess such spectral lines in sufficient numbers.  

Simulations were used to examine how extensive the averaging can be while simultaneously remaining valid, i.e., that the averaged line profile remains representative of a line whose strength and excitation potential are similar to that of the line-group average.  Figure~
\ref{fig:cool_dwarf_profiles} shows profiles from a model of a cool dwarf for three different line strengths at four different center-to-limb positions.  For a hotter model, profiles for five different line strengths were in Fig.\ 2 of Paper I, while Fig.\ 4 of that paper showed the excitation-potential dependence for different solar lines.  In all models -- at least for lines of moderate strength -- there is a continuous and smooth change of line shapes, such that the average of slightly stronger and slightly weaker line shapes (or such of slightly different excitation potentials) indeed closely agree with the profile of an intermediate-parameter line.

However, this smooth variation does not hold for the very strongest lines, which start to develop distinctive damping wings, and for such lines a separate treatment is called for.  As mentioned above, a few such very strong and also reasonably unblended lines were identified in the spectrum of Alopex but they were too few to permit averaging to sufficiently high S/N.   Also, the very weakest lines would merit a separate treatment, although their analyses will require lower noise spectra than presently available.

The range of line strengths for which such averaging is valid decreases for more subtle line parameters such as asymmetries and bisector shapes.  This was illustrated with synthetic bisectors in Figs.\ 5--7 of Paper~I.  Averaging will still be valid but has to be restricted to lines of a more limited parameter spread.  While not attempted in this work, searches for bisector dependence should become feasible with future data.  Also, averaging of lines from different species (\ion{Fe}{I}, \ion{Fe}{II}, \ion{Ti}{I}, \ion{Ti}{II}, etc.) would give access to a larger sample of photospheric lines with rather similar diagnostic potential, as seen in line-group averages from spectra of the Sun and Procyon \citep{dravins08}.

\begin{figure}[H]
\centering
  \includegraphics[width=8cm]{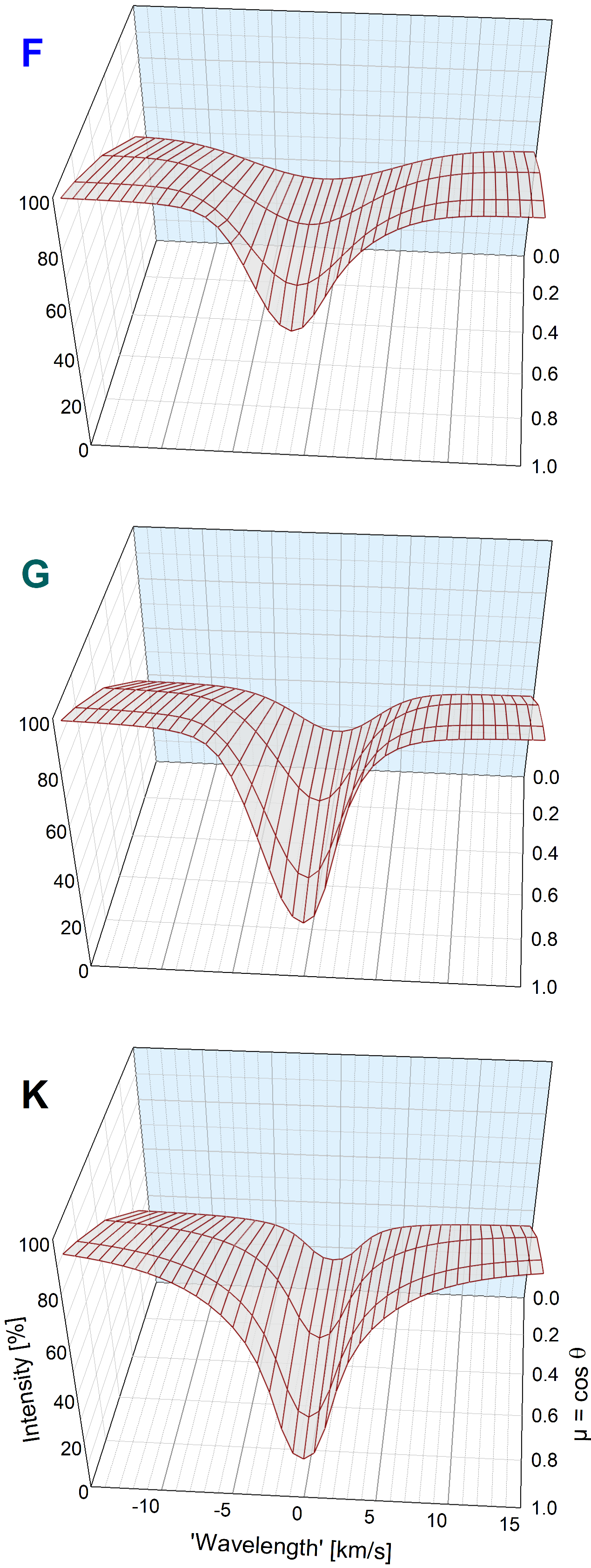}
     \caption{Profiles for an \ion{Fe}{I} line (the strongest one in Fig.\ \ref{fig:model_center_to_limb}) from models representing main-sequence stars of spectral types F, G, and K, with $T_{\textrm{eff}}$ = 6730, 5900, and 3960~K, show their quite different center-to-limb behavior.  The intensity scales are those of stellar disk centers and the varying continuum levels thus reflect the limb darkening, more pronounced in cooler stars. }
     \label{fig:3d_model_profiles}
\end{figure}

\subsection{Challenges in high-fidelity spectroscopy}

Clearly, these types of studies critically depend upon achieving the very best S/N levels.  Even if many recordings can (and must) be averaged, such  stacking typically only reduces the random noise as the square root of the number of exposures, a slowly growing function.  For efficiently probing stellar surfaces, spectra with initial S/N ratios on the order of 500 or more will be required, such that are becoming available with high-performance instruments such as PEPSI on LBT \citep{strassmeieretal15, strassmeieretal18a, strassmeieretal18b} or ESPRESSO on the ESO VLT \citep{pepeetal14, gonzalesetal17} .    

Although various 3\mbox{-}D atmospheric signatures are potentially observable in spatially resolved stellar spectra, achievable S/N limit the more realistically measurable quantities to the center-to-limb changes of entire line profiles, such as their overall widths and depths.  As illustrated in Fig.\ \ref{fig:model_center_to_limb}, even such quantities may strongly depend on the surface granulation structure, and the center-to-limb variation in line-widths are predicted to be drastically greater in hotter F-type stars than the cooler K-types.  Fig.\ \ref{fig:3d_model_profiles} shows these line-profile changes in more detail.  To verify such predictions requires observations of some suitably bright stars with large transiting planets.  The two planet host stars now analyzed in Papers~II and III were selected for their apparent brightness and planet size (Fig.\ 1 in Paper~II).  However, ongoing searches for bright host stars with transiting exoplanets keep finding new targets, a recent discovery being a hot Jupiter transiting the m$_{\textrm{V}}$ = 7.6, rapidly rotating A2~V star HD\,185603 with an 0.81\% transit depth \citep{lundetal17}.  Such a  $T_{\textrm{eff}}$ $\sim$8730~K star may have fewer measurable lines and pose other challenges but is just one example of a target for which spatially resolved stellar spectroscopy could explore new parameter domains of stellar astrophysics.

\begin{acknowledgements}
{This study used data obtained from the ESO Science Archive Facility, originating from observations made with ESO Telescopes at the La Silla Paranal Observatory under the two programs ID:079.C-0127(A) and ID:079.C-0828(A). The first was a Guaranteed Time Observation program by the HARPS Consortium (M.Mayor et al.): `A high-precision systematic search for exoplanets in the Southern Sky,' while the second was `Search for sodium in the atmosphere of the extrasolar planet HD189733b' by Lecavelier des {\'E}tangs et al.  HGL acknowledges financial support by the Sonderforschungsbereich SFB881 `The Milky Way System' (subproject A4) of the German Research Foundation (DFG).   The work by DD was performed in part at the Aspen Center for Physics, which is supported by National Science Foundation grant PHY-1066293.  DD also acknowledges stimulating stays as a Scientific Visitor at the European Southern Observatory in Santiago de Chile.  We thank the referee for detailed and insightful comments.}

\end{acknowledgements}

%\normalsize 

\begin{appendix}

\section{\bf{Listing of lines}}

{List of the 158 largely unblended \ion{Fe}{I} lines selected; 86 weaker lines (measured absorption-line central intensity $\geq$55\% of continuum) and 72 stronger lines.  Wavelengths and excitation potentials are from \citet{stenflolindegren77}; line-center intensities are in units of the normalized local continuum as obtained from averaged Gaussian fits to all current HARPS exposures.  These intensities are measured values, and have been subject to convolution with the spectrometer instrumental profile; observations with higher spectral resolution would yield deeper lines.  A graphical display of these sets of lines is in Fig.\ \ref{fig:selected_lines}. }

\begin{table*}
\caption{Selected weaker \ion{Fe}{I} lines}             
\label{table_weak_lines}      
\centering          
\begin{tabular}{c c c c c c c c c}
\hline\hline       
Wavel.\ [nm] & Int.\ [\%] & Exc.\ pot. [eV] & Wavel.\ [nm] & Int.\ [\%]  & Exc.\ pot. [eV] & Wavel.\ [nm] & Int.\ [\%] & Exc.\ pot. [eV] 
 \\ 
\hline                    
455.16499 & 62 & 3.94 & 553.92824 & 79 & 3.64 & 616.53641 & 61 & 4.14 \\
459.35268 & 61 & 3.94 & 555.26919 & 91 & 4.95 & 622.67403 & 71 & 3.88 \\
465.75879 & 55 & 2.84 & 556.88689 & 86 & 3.63 & 631.15050 & 68 & 2.83 \\
465.82976 & 72 & 3.27 & 560.76668 & 83 & 4.15 & 631.58164 & 63 & 4.07 \\
467.28364 & 59 & 1.61 & 561.13587 & 86 & 3.63 & 638.07483 & 59 & 4.19 \\
472.61396 & 73 & 3.00 & 561.86360 & 55 & 4.21 & 638.57206 & 88 & 4.73 \\
478.59583 & 67 & 4.14 & 563.58247 & 65 & 4.26 & 639.25429 & 73 & 2.28 \\
479.43571 & 76 & 2.42 & 563.66992 & 74 & 3.64 & 643.64102 & 88 & 4.19 \\
479.94092 & 58 & 3.64 & 564.66854 & 90 & 4.26 & 649.64738 & 57 & 4.79 \\
480.25216 & 81 & 4.61 & 565.14716 & 79 & 4.47 & 657.42325 & 57 & 0.99 \\
480.81509 & 60 & 3.25 & 565.23194 & 72 & 4.26 & 658.12143 & 71 & 1.48 \\
480.99400 & 70 & 3.57 & 566.13480 & 75 & 4.28 & 660.80301 & 74 & 2.28 \\
521.38071 & 88 & 3.94 & 567.76875 & 90 & 4.10 & 662.75488 & 76 & 4.55 \\
522.31851 & 63 & 3.63 & 568.02441 & 86 & 4.19 & 664.69355 & 85 & 2.61 \\
523.62039 & 65 & 4.19 & 572.08950 & 83 & 4.55 & 670.35720 & 62 & 2.76 \\
529.39609 & 67 & 4.14 & 579.39178 & 66 & 4.22 & 671.03213 & 74 & 1.48 \\
529.45493 & 79 & 3.64 & 580.77868 & 89 & 3.29 & 673.79830 & 82 & 4.56 \\
535.81168 & 84 & 3.30 & 581.19172 & 86 & 4.14 & 673.95243 & 78 & 1.56 \\
538.63345 & 64 & 4.15 & 581.48092 & 75 & 4.28 & 674.59626 & 92 & 4.07 \\
539.52187 & 77 & 4.44 & 582.78794 & 84 & 3.28 & 674.69508 & 92 & 2.61 \\
541.70405 & 65 & 4.41 & 583.51018 & 84 & 4.26 & 675.27107 & 74 & 4.64 \\
546.08762 & 88 & 3.07 & 585.22222 & 64 & 4.55 & 679.32656 & 87 & 4.07 \\
546.15530 & 71 & 4.44 & 585.87840 & 85 & 4.22 & 680.68491 & 64 & 2.73 \\
546.42825 & 63 & 4.14 & 586.11123 & 90 & 4.28 & 682.03730 & 69 & 4.64 \\
547.00957 & 71 & 4.44 & 588.12822 & 84 & 4.61 & 682.85976 & 61 & 4.64 \\
548.71489 & 67 & 4.41 & 590.24755 & 85 & 4.59 & 683.98340 & 67 & 2.56 \\
549.18346 & 85 & 4.19 & 592.96802 & 65 & 4.55 & 684.13450 & 60 & 4.61 \\
549.35012 & 65 & 4.10 & 603.40365 & 90 & 4.31 & 684.36606 & 59 & 4.55 \\
552.24491 & 58 & 4.21 & 608.27147 & 59 & 2.22  \\
\hline                  
\end{tabular}
\end{table*}

\begin{table*}
\caption{Selected stronger \ion{Fe}{I} lines}             
\label{table_strong_lines}      
\centering          
\begin{tabular}{c c c c c c c c c}
\hline\hline       
Wavel.\ [nm] & Int.\ [\%] & Exc.\ pot. [eV] & Wavel.\ [nm] & Int.\ [\%]  & Exc.\ pot. [eV] & Wavel.\ [nm] & Int.\ [\%] & Exc.\ pot. [eV] 
 \\ 
\hline                    
436.59004 & 37 & 2.99 & 525.06527 & 26 & 2.20 & 602.70562 & 49 & 4.07 \\
443.25726 & 39 & 3.57 & 525.34693 & 36 & 3.28 & 605.60114 & 49 & 4.73 \\
444.54760 & 33 & 0.09 & 536.54063 & 38 & 3.57 & 606.54921 & 29 & 2.61 \\
452.34015 & 47 & 3.65 & 536.74755 & 29 & 4.41 & 615.16217 & 50 & 2.18 \\
457.42191 & 48 & 3.21 & 536.99702 & 30 & 4.37 & 615.77331 & 52 & 4.07 \\
460.20006 & 24 & 1.61 & 537.37136 & 48 & 4.47 & 617.33433 & 43 & 2.22 \\
467.88519 & 23 & 3.60 & 537.95796 & 44 & 3.69 & 618.02084 & 51 & 2.73 \\
474.15341 & 29 & 2.83 & 538.94866 & 38 & 4.41 & 620.03204 & 41 & 2.61 \\
483.95500 & 39 & 3.27 & 543.29525 & 46 & 4.44 & 621.34375 & 36 & 2.22 \\
487.43565 & 64 & 3.07 & 544.50502 & 34 & 4.39 & 621.92886 & 35 & 2.20 \\
487.58815 & 40 & 3.33 & 547.39076 & 40 & 4.15 & 623.26493 & 43 & 3.65 \\
488.54361 & 35 & 3.88 & 552.55472 & 53 & 4.23 & 626.51412 & 35 & 2.18 \\
489.28624 & 50 & 4.22 & 554.39399 & 49 & 4.22 & 627.02322 & 52 & 2.86 \\
490.93874 & 42 & 3.93 & 563.82675 & 41 & 4.22 & 629.78013 & 40 & 2.22 \\
491.80152 & 51 & 4.23 & 566.25233 & 37 & 4.18 & 632.26936 & 41 & 2.59 \\
494.63941 & 26 & 3.37 & 570.15527 & 34 & 2.56 & 633.53378 & 33 & 2.20 \\
495.01108 & 34 & 3.42 & 570.54677 & 63 & 4.30 & 640.80262 & 42 & 3.69 \\
498.32566 & 31 & 4.15 & 571.78379 & 51 & 4.28 & 643.08538 & 32 & 2.18 \\
504.42164 & 34 & 2.85 & 573.17666 & 51 & 4.26 & 648.18784 & 46 & 2.28 \\
509.07807 & 35 & 4.26 & 575.31287 & 43 & 4.26 & 649.89461 & 49 & 0.96 \\
514.17460 & 26 & 2.42 & 577.50849 & 50 & 4.22 & 659.38798 & 39 & 2.43 \\
524.24988 & 31 & 3.63 & 591.62535 & 47 & 2.45 & 660.91189 & 47 & 2.56 \\
524.37823 & 45 & 4.26 & 595.66997 & 41 & 0.86 & 675.01597 & 43 & 2.42 \\
525.02171 & 29 & 0.12 & 600.79656 & 54 & 4.65 & 685.51684 & 54 & 4.56 \\
\hline                  
\end{tabular}
\end{table*}

\end{appendix}

\end{document}